\documentclass[nofootinbib,floats,aps,groupedaddress,preprint]{revtex4}
\usepackage{amssymb,amsmath,amsbsy}
\usepackage{mathrsfs}
\usepackage[dvips]{graphics}
\usepackage{epsfig}
\newenvironment{Eqnarray}%
     {\arraycolsep 0.14em\begin{eqnarray}}{\end{eqnarray}}

\def\beq{\begin{equation}}
\def\eeq{\end{equation}}
\def\beqa{\begin{Eqnarray}}
\def\eeqa{\end{Eqnarray}}
\def\ifmath#1{\relax\ifmmode #1\else $#1$\fi}
\def\lsup#1{^{\lower 4pt\hbox{$\scriptstyle#1$}}}
\def\llsup#1{^{\lower 2pt\hbox{$\scriptstyle#1$}}}

\def\lsim{\mathrel{\raise.3ex\hbox{$<$\kern-.75em\lower1ex\hbox{$\sim$}}}}
\def\gsim{\mathrel{\raise.3ex\hbox{$>$\kern-.75em\lower1ex\hbox{$\sim$}}}}

\def\eq#1{eq.~(\ref{#1})}

\def\Eq#1{Eq.~(\ref{#1})}

\begin{document}
\preprint{
\vbox{\vspace*{2cm}
      \hbox{arXiV:0904.4485 [hep-ph]}
      \hbox{April, 2009}
}}
\vspace*{3cm}

\title{Gauge Mediation with a small $\mu$ term and light squarks}


\author{John D. Mason}
\email[]{jdmason@physics.harvard.edu}
\affiliation{Jefferson Physical Laboratory, Harvard University\\
 Cambridge, MA 02138, U.S.A. \\
\vspace{2cm}}



\begin{abstract}

We consider a solution to the $\mu$ problem in the context of Non-Minimal Gauge Mediation with two Singlets and Low-Scale Messengers. This solution reduces tuning associated with the ``Little Hierarchy" problem by permitting a naturally small $\mu$ term, $\mathcal{O}(100-300 ~{\rm GeV})$, due to small mixing between the Singlets.  The smallness of $\mu$ also relies crucially on compressing the Gauge Mediated sparticle spectrum resulting in 330-400 GeV squarks. In addition to a small $\mu$ term, the theory achieves $m_{Higgs} > 114.4~{\rm GeV}$ through a large Higgs quartic coupling when $\tan{\beta} \sim 1.5$. The vacua studied are globally stable with all couplings perturbative to the GUT scale. The amount of tuning required to get the correct Electroweak scale is  $\mathcal{O}(10\%)$, with a similar residual tuning associated with the region of parameter space where the lightest CP-even Higgs mass is above the LEP bound. 
\end{abstract}


\maketitle

\section{Introduction}

Supersymmetry (SUSY) can stabilize the hierarchy between the Planck ($M_p$) and Electroweak ($M_{ew}$) scales. If SUSY is realized in Nature, it is a broken symmetry, and can be described by The Minimal Supersymmetric Standard Model (MSSM). In addition to supersymmetric versions of Standard Model interactions, the MSSM langrangian contains explicit SUSY breaking operators  ($\mathcal{L}^{soft}$), which break SUSY softly by relevant operators with a mass scale $\mathcal{O}(M_{ew})$.  Ultimately, the soft SUSY breaking lagrangian is assumed to originate from a model of Dynamical Supersymmetry Breaking (DSB) that spontaneously breaks supersymmetry at small scales using a supersymmetric form of Dimensional Transmutation. After generating a small vacuum expectation value (VEV) for an auxiliary component of some superfield ($X$), this field can then couple to the MSSM fields via Gauge Mediation and generate $\mathcal{L}^{soft}$  \cite{Dine:1981za,Dimopoulos:1981au,Dine:1981gu,Nappi:1982hm,AlvarezGaume:1981wy,Dimopoulos:1982gm} and more recently, \cite{Dine:1993yw,Dine:1994vc,Dine:1995ag} (for a review see \cite{Giudice:1998bp}). Models of Gauge Mediation are appealing because the scalar masses are positive and the same order of magnitude as the gaugino masses, provide a natural framework for minimal flavor violation, and are typically paramaterized in terms of a few number of parameters. 

The simplest model of Gauge Mediation is one in which the Messengers couple to a spurion, $X$, through the Yukawa interaction
\beq \label{messyuk}
W= \sum_i^N \lambda' X \tilde{\phi}^i\phi^i,
\eeq
where $N$ is the number of vector-like pairs of Messengers. This simple model is that of ``Minimal Gauge Mediation" (MGM). When $X= \langle X \rangle -\theta^2 F$, MGM determines all of the squark and slepton masses at the Messenger threshold in terms of only one parameter, $\Lambda = \frac{F}{\langle X \rangle}$, at leading order in $\frac{F}{\langle X \rangle^2}$.

Despite its success, models of MGM encounter two main problems. The first is the ``$\mu$ problem."  A superpotential operator of the form 
\beq \label{muterm}
W = \mu H_DH_U
\eeq
is required both to generate the correct Electroweak scale through the Electroweak minimization condition
 \beq \label{ewmin}
 \mu^2 = \frac{m_{H_D}^2 - m_{H_U}^2\tan^2{\beta}}{\tan^2{\beta}-1} -\frac{m_Z^2}{2},
 \eeq
as well as to lift the lightest Chargino above the experimental bound of $103 ~{\rm GeV}$ \cite{Abdallah:2003xe,Abbiendi:2003sc}. Since \eq{muterm} is a supersymmetric interaction its natural order of magnitude is $\mathcal{O}(M_p)$. Thus the fact that $\mu \sim M_{ew}$ is either a coincidence of mass scales, or the origin of $\mu$ is connected to the dynamics that generates $\mathcal{L}^{soft}$. The latter explanation is the most natural but requires an extension of the MGM model. There are a variety of proposals that solve the $\mu$ problem in MGM \cite{Dvali:1996cu,Chacko:2001km,Delgado:2007rz,Murayama:2007ge,Roy:2007nz,Giudice:2007ca}. 

The second problem that arises in MGM is the ``Little Hierarchy" problem. A broad class of SUSY models have a ``Little Hierarchy" problem due to the the LEP mass bound ($m_{h^0}>114.4 ~{\rm GeV}$) for a Standard Model Higgs \cite{Schael:2006cr}. In models where the lightest CP-even Higgs has a Standard Model-like coupling to the Z-boson and Standard Model-like decays, the tree-level Higgs mass is in conflict with the experimental bound.  Heavy squark masses can radiatively lift the lightest CP-even Higgs mass above the LEP bound \cite{Haber:1990aw}. However, heavy squarks also renormalize the up-type soft Higgs mass as
\beq \label{hsoftmass}
\delta m_{H_U}^2 = -{3 y_t^2\over 4 \pi^2} m_{\tilde t}^2 \ln(\lambda'\langle X \rangle/m_{\tilde t})< -(600 {\rm GeV})^2.
\eeq
In order to satisfy \eq{ewmin}, one typically must tune either the value of $\mu$ or the threshold value of $m_{H_U}^2$ in order to get the correct value for $m_Z$. 
The situation in MGM is somewhat worse because in MGM a ``Little Hierarchy" exists independent of the LEP Higgs mass bound (see \cite{Dine:2009gy} for a review).  The ``Little Hierarchy" in MGM is due to the right-handed slepton mass bound ($m_{e_R} > 73~{\rm GeV}$) \cite{PDBook}. This mass bound in combination with the MGM relation that $\frac{m_{stop}}{m_{e_R}} \sim \frac{g_3^2}{g_1^2}$ requires $m_{stop} > 750~{\rm GeV}$.  Large stop masses again induce the radiative correction in \eq{hsoftmass}, and \eq{ewmin} requires $\mu > ~ 600~{\rm GeV}$. A simple estimate of the degree of tuning required in \eq{ewmin} for MGM is 
\beq \label{tmeas}
T\sim \frac{\mu^2}{\frac{m_Z^2}{2}}.
\eeq
For $\mu  > (600 ~{\rm GeV})$, this induces a tuning in MGM of  $T >  89$, corresponding to at least a $1\%-2\%$ tuning for any theory of MGM. 

This paper will address both of these problems. Our goal is to construct a model with a dynamically generated $\mu$ term as small as possible while still having $m_{h^0} > ~114.4~{\rm GeV}$. From \eq{tmeas} it follows that a small value for $\mu$  can help reduce the ``Little Hierarchy."  From \eq{ewmin} and \eq{hsoftmass}, a stop mass as small as possible is required to do this without tuning and therefore a mediation mechanism different from that of MGM is required. We will refer to such models as Non-Minimal Gauge Mediated Models (N-MGM). Original models of DSB were in fact of the Non-Minimal variety \cite{Izawa:1997gs,Dimopoulos:1996yq}. In \cite{Martin:1996zb}, generalized Messengers were shown to yield a Non-Minimal spectrum, and in \cite{Agashe:1997kn}, models with doublet-triplet splitting in the Messenger sector were used to generate a Non-Minimal spectrum. More recently, a Non-Minimal spectrum in the broader context of DSB was discussed in \cite{Dine:2007dz}. In \cite{Cheung:2007es, Buican:2008ws, Abel:2008gv} a Non-Minimal spectrum was shown to emerge from theories of Direct Mediation.  In \cite{Meade:2008wd} it was shown that even the most general Non-Minimal Gauge Mediated spectrum, so-called ``General Gauge Mediation," has a predictive spectrum. Model building challenges and simple models of General Gauge Mediation were discussed in \cite{Carpenter:2008wi,Carpenter:2008rj,Buican:2008ws}. Following these results, we will consider a light squark scenario by utilizing a simple two-parameter realization of N-MGM. 

In \cite{Komargodski:2008ax} it was shown that the framework of General Gauge Mediation can be extended to provide a general description of the $\mu$ problem and its solutions. Solutions to the $\mu$ problem were shown to fall into two classes. In one class of models the $\mu$ term is generated via operators of the form: $ H_U\mathcal{O}_U + H_D \mathcal{O}_D$. A particularly nice realization of this class is \cite{Csaki:2008sr}, where a small $\mu$ term arises despite having stop masses of order $\mathcal{O}(1 ~ {\rm TeV})$. In this paper we will focus on the second class of models, those using operators of the form $SH_DH_U$ to generate the $\mu$ term. In \cite{Liu:2008pa}, a consistent theory of this type was found to work when Renormalization Group Evolution generates the various soft parameters of the NMSSM. In that case, the SUSY spectrum is heavy, $\tan{\beta}$ is large, and consistent vacuum solutions have a small $\mu$ term. We will consider the same two-parameter model in a different context and will find a qualitatively different phenomenology (e.g. a light sparticle spectrum), though we find a small $\mu$ term as well. For this reason, we consider our discussion complementary to those in \cite{Csaki:2008sr} and \cite{Liu:2008pa}.

While lowering the squark mass can alleviate the ``Little Hierarchy" problem, it removes the mechanism that lifts the lightest CP-even Higgs mass above the LEP bound. There are ways in which such models remain phenomenologically viable. The first is to assume that the lightest CP-even Higgs in the MSSM is not Standard Model-like \cite{Dermisek:2005ar,Chang:2005ht,Dermisek:2006wr,Chang:2008cw}. The second method is to keep the lightest CP-even Higgs Standard Model-like but replace the large Higgs quartic couplings that were induced, in MGM, by radiative corrections from heavy stop loops \cite{Haber:1990aw,Haber:1993an}, with some other new physics at the TeV scale that increases the Higgs quartic couplings \cite{Brignole:2003cm,Dine:2007xi}. In this paper we construct a realization of the latter option. Furthermore, as we will show, the mechanism that generates the Higgs quartic coupling in a light stop scenario will be connected to the  solution to the $\mu$ problem. 

The operator that can both increase the Higgs quartic coupling as well as solve the $\mu$ problem is well known, and it appears in the NMSSM \cite{Ellis:1988er}. In the Singlet extended MSSM (NMSSM), an R-symmetry (or a PQ symmetry) can forbid a bare $\mu$ term and allow the interaction
\beq \label{nmssm}
W = \lambda N H_D H_U.
\eeq
If the Singlet gets a large SUSY breaking mass, this operator can increase the Higgs quartic coupling \cite{Espinosa:1991wt, Espinosa:1991gr, Nomura:2005rk}. Formally, the Singlet can be integrated out generating the non-decoupling Higgs quartic coupling
\beq \label{quartic}
V = \lambda^2 |H_DH_U|^2.
\eeq
When  $\tan{\beta} \sim 1$ and $\lambda \sim .7$, this can boost the tree-level Higgs mass above the LEP bound.  In \cite{Batra:2004vc} this was shown to have significant effects if $\lambda$ is especially large.  We will use this mechanism to lift the physical Higgs mass, but we will not formally integrate out the Singlet since it will only have a mass of $1.5 ~{\rm TeV}$ in our case. 

The operator in \eq{nmssm} can also solve the $\mu$ problem if the scalar and F-term components of the Singlet get VEVs 
\beq 
N = \langle N \rangle - \theta^2 F_N
\eeq
and/or an A-term is generated 
\beq 
V =-\lambda A_{\lambda}NH_DH_U + {\rm h.c.}
\eeq
The $\mu$ problem can be solved with 
\beq 
\mu = \lambda \langle N \rangle,
\eeq
\beq
B\mu = \lambda A_{\lambda} \langle N \rangle -\lambda F_N,
\eeq
where $B\mu$ enters the Higgs potential as 
\beq
V= -B\mu H_DH_U + {\rm h.c.}
\eeq
We have used notation where $H_DH_U = \epsilon_{ij}H_D^iH_U^j = H_D^0H_U^0-H^-H^+$. Notice that this can provide an origin to the $\mu$ term while generating a $B\mu$ term of the same order of magnitude as long as $ |F_N| \sim |\mu|^2$ and/or $|A_{\lambda}| \sim |\mu|$. Many simple mechanisms that generate $\mu$ at the appropriate order of magnitude generate a $B\mu$ term that is too large to allow for stable Electroweak Symmetry Breaking \cite{Giudice:1998bp}. In this way the $\mu$ problem is also associated with the problem of the origin of $B\mu$, and is also referred to as the $\mu/B\mu$ problem. 

While the operator in \eq{nmssm} can lift the Higgs mass and solve the $\mu$ problem, it is challenging to accomplish both of these tasks simultaneously. Typically one must give up perturbativity of $\lambda$ to the GUT scale or resort to having large stop masses. The reason for this is as follows. In order to solve the $\mu$ problem, the Singlet field N must get a scalar VEV. However,  in order to generate the non-decoupling quartic interaction in \eq{quartic}, the propagating degree of freedom must be heavy. In theories where a Singlet's VEV is driven by a balance between its quartic coupling ($\kappa^2$)  and a negative soft mass-squared, the mass of the propagating Singlet ($m_n$) is related to the VEV ($\langle N \rangle$) via the relation 
\beq \label{singmassrel}
m_n \sim \kappa \langle N \rangle.
\eeq
If $\kappa < 1$, as is typical in a theory with a Singlet that is perturbative to the GUT scale, then one cannot decouple the propagating Singlet without also increasing $\mu$ (note that we must keep $\lambda$ large and fixed or else we would lose the boost in the Higgs quartic coupling). From \eq{tmeas} we see that this would increase the ``Little Hierarchy" problem. Furthermore, in a theory with large $\mu$ and fixed $\lambda$, the lightest CP-even Higgs is Standard Model-like and has a mass given by \cite{Ellis:1988er}:
\beq \label{nmssmhiggs}
m^2_{h^0} \sim m_Z^2\cos^2{2\beta} + \lambda^2v^2\sin^2{2\beta} -\frac{\lambda^2v^2}{\kappa^2}(\lambda-\kappa \sin{2\beta})^2.
\eeq
The additional negative contribution to $m^2_{h^0}$ is present because of mixing with the Singlet. Despite the fact that the Singlet is formally decoupled in this limit, this non-decoupling contribution to the lightest CP-even Higgs mass remains.  If we require that $\kappa$ and $\lambda$ are perturbative to the GUT scale and fix $\tan{\beta} \sim 1.5$, one still requires $m_{sq} > 600~ {\rm GeV}$ in order to agree with the LEP Higgs bound or allow coupling non-perturbative before the GUT scale. For this choice of $m_{sq}$ one finds $\mu > 550 ~{\rm GeV}$ and therefore at least a $1\%$ tuning. In actual implementations of the NMSSM tunings are worse. So we see that while in principal adding a Singlet to the MSSM could solve both the $\mu$ problem and lift the Higgs mass above the LEP bound with light squarks, in practice, one still relies on heavy stops to lift the Higgs mass because of the mixing effect. In this way, the ``Little Hierarchy" is re-introduced into NMSSM scenarios that solve the $\mu$ problem. Said another way, in a single Singlet model, there is a conflict between getting a small $\mu$ term and raising the Higgs mass with the operator in \eq{nmssm}.

One way around this problem, invoked in \cite{Barbieri:2007tu,Cavicchia:2008fn}, is to allow the Singlet VEV to be driven by a large $A$-term ($A_{\lambda} \sim 300~ {\rm GeV}$) rather than a negative $m_N^2$.  Small mixing can be achieved when the the Singlet soft mass is small and positive. However, since the soft mass is not protected by any symmetry, the requirement that $m_N < 50~ {\rm{GeV}}$ in such models remains a model building challenge.

In this paper we will construct models of Non-Minimal Gauge Mediation that use the operator of \eq{nmssm} to \emph{simultaneously} lift the lightest CP-even Higgs mass above the LEP bound as well as generate a small $\mu$ term, breaking Electroweak symmetry in a globally stable vacuum. The models we will consider have two Singlets, in addition to the Gauge Mediated MSSM, and allow the Singlets to couple directly to the Messengers of Gauge Mediation.  After integrating out the Messengers, the Singlets will have soft masses and various trilinear couplings.  For certain choices of Messenger-Singlet couplings, these two Singlets can mix such that the Singlet appearing in \eq{nmssm} avoids the typical NMSSM relation, \eq{singmassrel}, between its mass and VEV. This will permit the Singlet to have a large mass and a large coupling ($\lambda$) but a small VEV. As a result, a small $\mu$-term can be generated along with a sizable $B\mu$ term from the soft trilinear coupling. This will satisfy the Electroweak minimization condition, \eq{ewmin}, only if the squarks are not very heavy. Conveniently, in these models, heavy squarks are not required to make the Higgs mass heavier than the LEP bound due to an additional contribution to the Higgs quartic coupling from the heavy Singlet equation of motion. This new quartic interaction only raisies the Higgs mass above the LEP bound for $\tan{\beta} \sim 1$; we will show that a sufficiently large $B\mu$ (and thus sufficienly small $\tan{\beta}$) can be achieved.  Ultimately, these models provide a mechanism to generate small value for $\mu$ reducing the tuning required to get the correct Electroweak scale and thus reducing the ``little hierarchy" problem.  

In section II we outline the basic mechanism that allows the operator in \eq{nmssm} to simultaneously solve the $\mu$ problem and lift the Higgs mass without the need for heavy stops. In section III we will review two different two parameter models of Non-Minimal Gauge Mediation, which will allow us to realize a light stop scenario without violating model independent experimental bounds of sparticle searches. In one model we will impose GUT relations between doublet and triplet Messenger couplings. In a second model we will relax this assumption. In section IV we will review the use of direct Messenger-Singlet couplings in the NMSSM both in the context of MGM and the two parameter models of Non-Minimal Gauge Mediation. Finally in section V we will extend the NMSSM to include an additional Singlet and show that this theory naturally realizes the mechanism discussed in Section II, which uses the operator in \eq{nmssm} to lift the Higgs mass and solve the $\mu$ problem.  We will investigate the phenomenology of our two models, evaluate the phenomenological parameters $\tan{\beta}$ and $\mu$ when $m_{h^0} > 114.4~{\rm GeV}$, and show that $\mu$ can be very small in these models. Finally we will comment on the size of the phenomenologically allowed region of parameter space in each of these models showing that there is some mild tuning.

\section{Small $\mu$ and Large quartic coupling via mixing}

In Section I, we argued that \eq{nmssm} cannot be used to simultaneously increase the Higgs quartic coupling and yield a small value for $\mu$. The obstruction to this is the simple fact that for the Singlet of the NMSSM with Superpotential interactions given by 
\beq \label{nmssmn3}
W=\lambda N H_D H_U -\frac{\kappa}{3}N^3,
\eeq
and soft scalar potential given by 
\beq
V^{(soft)}=m_N^2|N|^2-\lambda A_{\lambda}NH_DH_U -\frac{\kappa}{3}A_{\kappa}N^3 + \rm{h.c.},
\eeq
if N gets a VEV due to a negative soft mass
\beq
N|_{\theta=0}= \langle N \rangle + n,
\eeq
the propagating field ($n$) gets mass smaller than the VEV, and cannot decouple the mixing effect in \eq{nmssmhiggs} without also decoupling $\mu$ and re-introducing a ``Little Hierarchy."  In order to overcome this challenge, one requires that
\beq \label{massrel}
\lambda \langle N \rangle < m_n.
\eeq 
Taking $\lambda$ small would appear to help, but later we will find larger values of $\lambda$ are required to satisfy the LEP Higgs mass bound. For this reason we will fix $\lambda \sim (0.65, 0.7)$ for the remainder of the paper. Then there are only two ways in which \eq{massrel} can be achieved. The first is to allow a large value for the Singlet quartic coupling ($\kappa$), because a tree-level relation is that $m_n \sim \mu\frac{\kappa}{\lambda}$. Since we will restrict our attention to theories that are perturbative to the GUT scale, we will not consider $\kappa > .63$. So we will not consider this first option. The second option is to allow the dynamics of mixing to permit a relation like \eq{massrel}.  Such a relation arises in the Higgs Sector of the MSSM in the decoupling limit when $\tan{\beta}$ is large. It is useful to review this limit. The VEVs of the Higgs fields are 
\beqa
H_U = \left( \begin{array}{c} 0  \\
v_u
\end{array}
\right)\, ~~~
H_D = \left( \begin{array}{c} v_d  \\
0
\end{array}
\right).\,
\eeqa
In the decoupling ($m_A \gg m_Z$) and large $\tan{\beta}$ limit, where $\tan{\beta} >> \frac{m_A^2}{m_Z^2}$, the CP-even Higgs mass matrix in the basis $(H_D, H_U)$ is 
\beqa
M_{CPE}  \sim \left( \begin{array}{cc} m_A^2  & 0 \\
0&m_Z^2 
\end{array}
\right).\,
\eeqa
In this case, the neutral CP-even component of $H_D$ is essentially a heavy mass eigenstate with a  VEV ($v_d$) and a mass ($M_A$) satisfying the relation: $v_d \ll M_A$.

We can achieve a similar result for the NMSSM Singlet, but first we need to extend the NMSSM to contain an additional Singlet (``S") in order to get the desired mixing. Ignoring the MSSM Higgs doublets, the superpotential and scalar potential of a two-Singlet extended MSSM can have the form
\beq
W= \lambda N H_DH_U -\frac{\kappa}{3}S^3,
\eeq
 \beq \label{VSN}
V^{(scalar)}=m_N^2|N|^2+m_S^2|S|^2+[bSN +\rm{h.c.}]+\kappa^2|S|^4.
\eeq
Note that since $S$ and $N$ are Singlets, we must explain why many interactions that \emph{a priori} are not forbidden have not been written down. Later this will arise due to the presence of an $R$-symmetry.

If $b^2 > m_N^2m_S^2$, then the origin will be destabilized and the $S$ and $N$ fields will get vacuum expectation values. Defining 
\beq \label{betaprime}
\tan{\beta'} = \frac{\langle S \rangle }{\langle N \rangle} 
\eeq
and expanding the potential about the VEVs, one finds that the CP-even mass matrix for these fields in the basis $(N,S)$ is
\beqa \label{simplemodel}
M_{CPE}  \sim \left( \begin{array}{cc} -b\tan{\beta'}  & b \\
b&-b\cot{\beta'} +  4\kappa^2s^2
\end{array}
\right).\,
\eeqa
Now, if 
\beq \label{rels}
m_N^2 \gg b \gg  m_S^2 > 0,
\eeq
then $\tan{\beta'} \gg 1$ and the mixing between the $N$ and $S$ states is small. In this case, the field $N$ has a VEV smaller than the mass of its propagating component:
\beq
\langle N \rangle \sim \frac{m_N }{\sqrt{\kappa} \tan{\beta'}^2}.
\eeq
If we couple $N$ to the Higgs via \eq{nmssm}, we get the desired result. We can generate a small $\mu$ term while at the same time decoupling the mixing effect of \eq{nmssmhiggs} that reduces the lightest CP-even Higgs mass. 

We will see now that the scalar potential in \eq{VSN} with the relations in \eq{rels} can arise in models of Non-Minimal Gauge Mediation.  First we review two parameter models of Non-Minimal Gauge Mediation. 

\section{Two Parameter Non-Minimal Gauge Mediation}

In order to realize a Gauge Mediated scenario where the squark masses are light ($m_{sq} \sim \mathcal{O}(350~ {\rm GeV})$) without violating the mass bound on the right-handed selectron, one must consider models that deviate from MGM. squark, slepton, and gaugino masses in MGM are 
\beq \label{ogmrel}
m_f^2= 2\sum_{r=1}^{3}C^r_f\left( \frac{\alpha^{(r)}}{4\pi} \right)^2\left| \frac{F}{X}\right|^2,~~~~M^{(r)}_{\lambda}=\frac{\alpha^{(r)}}{4\pi}\frac{F}{X}.
\eeq

For the scope of this paper, we focus on two different two-parameter extensions of MGM, because they will most simply illustrate the success of the two-Singlet mixing mechanism and eventually lead to the scalar potential in \eq{VSN}. The first two parameter model we consider assumes that the doublet and triplet Messengers come from the same representation of $SU(5)$. The second two parameter model relaxes this assumption and treats the doublet and triplet Messengers and their couplings independently. 

\subsection{N-MGM doublet and triplet Messengers from the same $SU(5)$ representations}

Consider a Hidden Sector with two Spurions that have both scalar VEVs and F-term VEVs. 
\beq
X_1 = \langle X_1 \rangle -\theta^2 F_{X_1}, ~~~X_2 = \langle X_2 \rangle -\theta^2 F_{X_2}.
\eeq
Now consider a model with a vector-like pairs of Messengers: ($\phi_1$, $\tilde{\phi_1}$) transforming as ${\bf 5} +{\bf \bar{5}}$ under $SU(5)$. Let these couple through the Yukawa interactions 
\beq \label{gut}
W= (\lambda_1X_1+\lambda_2X_2)\tilde{\phi_1}\phi_1.
\eeq
The generally complex couplings, $\lambda_1$ and $\lambda_2$, and parameters $F_{X_1}, F_{X_2}, \langle X_1 \rangle, \langle X_2 \rangle$,  will introduce CP-violating phases into the gaugino mases that cannot be rotated away by field redefinition. This will result in a contribution to CP-violating observables such as the electron and neutron electric dipole moments. In order to evade current experimental bounds on the electron EDM ($d_e < 10^{-27} e~ cm$) the physical phase ($\theta$) appearing in EDM computations must be $\theta < \mathcal{O}(10^{-2})$ \footnote{ More specifically this bound is $\theta < \mathcal{O}(\frac{10^{-2}}{\tan{\beta}})$.  Since the models of Section V will have $\tan{\beta} \sim 1$, requiring $\theta \leq 10^{-2}$ will be sufficient for our purposes. This level of tuning is unavoidable without specifying the underlying mechanism generating the Singlet VEVs; this is just the SUSY CP problem.  If the VEVs ($F_{X_1}, F_{X_2}, \langle X_1 \rangle, \langle X_2 \rangle$) are taken to be real, then the phases of $\lambda_1$ and $\lambda_2$ can be removed by field redefinition. However, if GUT relations between $\lambda_1$ and $\lambda_2$ are not assumed, then there will be physical CP-violating phases in the theory which are subject to the EDM constraints}.  Since the physical phase must be small we shall ignore it in the analysis that follows and define 
\beq \label{lambdas}
\Lambda_q = \frac{ \lambda^{(T)}_1F_{X_1} +\lambda^{(T)}_2F_{X_2} }{  \lambda^{(T)}_1\langle X_1 \rangle +\lambda^{(T)}_2 \langle X_2 \rangle},~~~~~
\Lambda_{\ell}= \frac{ \lambda^{(D)}_1F_{X_1} +\lambda^{(D)}_2F_{X_2} }{  \lambda^{(D)}_1\langle X_1 \rangle +\lambda^{(D)}_2 \langle X_2 \rangle},
\eeq
\beq
\Lambda_c^2 = \Lambda_q^2;~~\Lambda_w^2 = \Lambda_{\ell}^2;~~\Lambda_Y^2
= \left ({2 \over 3} \Lambda_q^2 + \Lambda_{\ell}^2 \right ),
\label{scalarsemgm}
\eeq
where $\lambda_i^{(T)}$ and $\lambda_i^{(D)}$ differ due to RG running from the GUT scale. The squark, slepton, and gaugino masses are given as
\beq
m_{g} = {\alpha_3 \over 4 \pi} \Lambda_q,~~~~
m_w = {\alpha_2 \over 4 \pi} \Lambda_{\ell},~~~~
m_b =  {\alpha_1 \over 4 \pi} \left [{2\over 3} \Lambda_q +
\Lambda_\ell \right ],
\label{gluinoformula}
\eeq
\beq
\label{squarkformula}
\widetilde m_f^2 = 2
\left[
C_3\left({\alpha_3 \over 4 \pi}\right)^2 \Lambda_c^2
+C_2\left({\alpha_2\over 4 \pi}\right)^2 \Lambda_w^2
+{5 \over 3}{\left(Y\over2\right)^2}
\left({\alpha_1\over 4 \pi}\right)^2 \left(\frac{2}{3}\Lambda_Y^2 +\Lambda_{\ell}^2 \right) \right].
\eeq
The squark, slepton, and gaugino spectrum is determined by only two parameters, $\Lambda_q$ and $\Lambda_{\ell}$.  Here we have given only the finite threshold values for the sparticle masses. In general there will be corrections due to the fact that these soft parameter run from the Messenger scale down to $M_{ew}$. Since we are working with Low-Scale Gauge Mediation, most of these RG effects may be ignored, though we will always include the RG running due to the top Yukawa, like those in \eq{hsoftmass}.

Now we can allow $\Lambda_q < \Lambda_{\ell}$ such that $\frac{m_{sq}}{m_{sl}}\sim 2$ rather than the typical MGM hierarchy $\frac{m_{sq}}{m_{sl}}\sim 10 $ . For instance, $\Lambda_q \sim 17.5~{\rm TeV}$ and $\Lambda_{\ell} \sim 85~{\rm TeV}$ allows $m_{sq} \sim 370 ~{\rm GeV}$ and $m_{sl} \sim 150 ~{\rm GeV}$ and satisfies basic model-independent experimental bounds of the right-handed Selectron \cite{PDBook}, Gluino \cite{Alwall:2008zz}, and lightest Chargino \cite{Abbiendi:2003sc}. The MSSM sparticle mass bounds become stronger when the particle in question is the NLSP.  This is of particular importance when considering models of Non-Minimal Gauge Mediation \cite{Carpenter:2008he,Rajaraman:2009ga}. In the model we will consider in Section V, none of the Standard Model partners will be the NLSP. Rather the NLSP will be a moderately light Singlino.

\subsection{N-MGM doublet and triplet Messengers from different $SU(5)$ representations}

If the doublet and triplet Messengers arise from different GUT multiplets, then the couplings of the Messenger bilinears $\tilde{\phi}^{(T)}\phi^{(T)}$ and $\tilde{\phi}^{(D)}\phi^{(D)}$ do not need to couple to fields with interaction strengths constrained to be equal at the GUT scale. In \cite{Agashe:1997kn},  similar scenarios were used to compress the sparticle spectrum as well as help generate a VEV for the Singlet N in the NMSSM. In what follows here, we will see that such models have a larger region of allowed parameter space and thus a reduced amount of tuning.

First consider the simplest model with only one Spurion:
\beq
W = \lambda^{(T)} X \tilde{\phi}^{(T)}\phi^{(T)} + \lambda^{(D)} X \tilde{\phi}^{(D)}\phi^{(D)}.
\eeq
In this case, despite the different couplings, the spectrum of gauginos, squarks, and sleptons is that of MGM with $N=1$. A simple way to get a Non-Minimal sparticle spectrum again requires two Spurions. Let us modify the superpotential with the most general interactions when two Spurion Singlets are present \footnote{One can assume that $X_1$ and $X_2$ are distinguished by some global charge which forbids $X_1$-triplet and $X_2$-doublet couplings. However we do not consider this for two reasons. The most common way to give F-term VEVs to Spurions is to include a superpotential coupling, $W=FX$. This determines that the Spurion is truly a Singlet. Second, in the models that follow it will be crucial that Messengers get SUSY Breaking masses from two different F-terms.}
\beq \label{nogut}
W=  (\lambda_1^{(T)}X_1+ \lambda_2^{(T)} X_2) \tilde{\phi}^{(T)}\phi^{(T)}+ (\lambda_1^{(D)} X_1+ \lambda_2^{(D)}X_2)  \tilde{\phi}^{(D)}\phi^{(D)}.
\eeq

The resulting spectrum is given as in \eq{lambdas}, but without GUT relations between $\lambda^{(T)}$ and $\lambda^{(D)}$.  Just as in the previous case one can choose $\Lambda_q < \Lambda_{\ell}$ and get a compressed sparticle spectrum.

We have realized in Low-Energy Non-Minimal Gauge Mediation, a compressed sparticle spectrum with light squarks and sleptons that can be heavier than the most basic model independent experimental bounds. Two main reasons attract our interest to these light squark scenarios. First, it is interesting to know if a realistic model of Gauge Mediation can accommodate such a spectrum. Secondly, such a spectum has the potential to relieve the ``Little Hierarchy" problem because this lessens the stop radiative corrections. Including renormalization due to the top Yukawa, the soft Higgs mass is now $ -(200~{\rm GeV})^2< m_{H_U}^2 < 0$ rather than the much larger MGM values of $-(600~{\rm GeV})^2$. This allows a naturally small value for $\mu$ and thus less of a tuning in order to get the correct Electroweak scale. Now we shall see this Non-Minimal framework also offers a new approach to the $\mu$ problem. 

\section{Direct Singlet-Messenger Couplings in MGM and N-MGM}

Extending MGM to include the interactions in \eq{nmssmn3} alone is not sufficient to solve the $\mu$ problem. This issue was discussed in \cite{deGouvea:1997cx}, generating a VEV for $N$ of the correct order of magnitude requires a substantial soft lagrangian for the Singlet $N$. In general this soft lagrangian is 
\beq \label{nmssm11}
V^{(soft)}=m_N^2|N|^2+[-\lambda A_{\lambda}NH_DH_U -\frac{\kappa}{3}A_{\kappa}N^3 + \rm{h.c.}].
\eeq
When only the interactions of \eq{nmssmn3} are included, the soft parameters, $m_N^2$, $A_{\lambda}$, and $A_{\kappa}$, are generated indirectly through the SUSY breaking felt by the Higgs fields. This in turn makes $m_N^2$ too small and requires a value for $\kappa$ that is too small to allow a locally stable minimum. The NMSSM as given by \eq{nmssmn3} must be augmented in MGM in order to generate a sizable soft lagrangian. One class of augmentations to the NMSSM one can use in MGM includes direct Singlet-Messenger couplings in order to generate a substantial soft lagrangian \cite{Han:1999jc,Ellwanger:2008py,Delgado:2007rz}.  We will now review a model of this class, \cite{Delgado:2007rz}, and the reason it fails to solve the $\mu$ problem if Gauge Mediation is Non-Minimal. 

In \cite{Delgado:2007rz}, the NMSSM in the context of MGM was extended to include direct Singlet-Messenger interactions,
\beq
W= X(\tilde{\phi_1}\phi_1 + \tilde{\phi_2}\phi_2) + \xi N \tilde{\phi_2}\phi_1.
\eeq
It was noticed  in \cite{Dvali:1996cu} that after integrating out the Messengers, the one-loop threshold contribution to the soft mass $m_N^2$ vanishes to leading order in $\frac{F}{X^2}$, and the two-loop value can be negative, driving a VEV for the Singlet N and the $\mu$ problem can be solved. It was also noticed that if the Messenger fields feel SUSY breaking by coupling to different Spurions through the interactions
\beq \label{sup1}
W=\lambda_1 X_1\tilde{\phi_1}\phi_1 +\lambda_2 X_2\tilde{\phi_2}\phi_2 + \xi N \tilde{\phi_2}\phi_1,
\eeq
the one-loop contribution to $m_N^2$ does not cancel and is in fact positive,
\beq \label{mn}
m^2_N= \frac{5\xi^2}{(16\pi^2)}(\Lambda_1 -\Lambda_2)^2\frac{x^2}{(1-x^2)^3}\left( (1+x^2)\log{x^2}+2(1-x^2) \right),
\eeq
where 
\beq
\Lambda_1 = \frac{F_{X_1}}{X_1}, ~~~~~\Lambda_2 = \frac{F_{X_2}}{X_2},~~~~~ x = \frac{\lambda_1X_1}{\lambda_2 X_2}.
\eeq
Similarly, the formulae for $A_{\lambda}$ and $A_{\kappa}$ are modified, but are still generated at one-loop. The MSSM spectrum that derives from \eq{sup1} is equivalent to that of MGM, this holds even after one allows for GUT breaking to enter into the renormalization of the couplings $\lambda_1$ and $\lambda_2$. 

Now consider the models of two parameter N-MGM in \eq{gut} or \eq {nogut} and include a second vector-like Messenger pair into the model. A superpotential that generates the N-MGM spectrum is 
\beq \label{sup2}
W=(\lambda_1 X_1+\lambda_2 X_2)\tilde{\phi_1}\phi_1 +(\kappa_1 Y_1+\kappa_2 Y_2)\tilde{\phi_2}\phi_2.
\eeq
If we add the coupling $ \xi N \tilde{\phi_2}\phi_1$  to \eq{sup2}, a one loop mass is generated from integrating out the Messenger fields just as in the theory of \eq{sup1}, however the MSSM spectrum will be that of Non-Minimal Gauge Mediation. We will absorb the couplings of the model into the definitions of the 4 parameters that determine the entire MSSM mass spectrum ($\Lambda^{(\kappa)}_q$, $\Lambda^{(\kappa)}_{\ell}$, $\Lambda^{(\lambda)}_q$, $\Lambda^{(\lambda)}_{\ell}$). Here $\Lambda^{(\lambda)}_q$ and $\Lambda^{(\lambda)}_{\ell}$ are the same as $\Lambda_q$ and $\Lambda_{\ell}$ in \eq{lambdas}. Due to the second Messenger pair, we have 
\beq
\Lambda^{(\kappa)}_q =  \frac{\kappa^{(T)}_1F_{Y_1} +\kappa^{(T)}_2F_{Y_2} }{  \kappa^{(T)}_1\langle Y_1 \rangle +\kappa^{(T)}_2 \langle Y_2 \rangle},~~~~~
\Lambda^{(\kappa)}_{\ell}= \frac{\kappa^{(D)}_1F_{Y_1} +\kappa^{(D)}_2F_{Y_2} }{  \kappa^{(D)}_1\langle Y_1 \rangle +\kappa^{(D)}_2 \langle Y_2 \rangle}.
\eeq
In terms of these parameters, $m_N^2$ is given as in \eq{mn} but with
\beq \label{ggmpar}
\xi^2(\Lambda_1-\Lambda_2)^2 \rightarrow \frac{3(\xi^T)^2}{5}(\Lambda^{(\kappa)}_{q} - \Lambda^{(\lambda)}_{q})^2 + \frac{2(\xi^D)^2}{5}(\Lambda^{(\kappa)}_{\ell} - \Lambda^{(\lambda)}_{\ell})^2.
\eeq
Since the Singlet $N$ has a positive mass squared generated at one loop, a model of Non-Minimal Gauge Mediation of the form \eq{gut} or \eq{nogut} will not generate a $\mu$ term since N does not get a VEV. For this reason, the Singlet-Messenger coupling solution to the $\mu$ problem in Non-Minimal Gauge Mediation not work in the NMSSM. 

\section{Non-Minimal Gauge Mediation with Two Singlets}

We now extend the NMSSM to include a second Singlet. We will see that inclusion of this field will permit a viable solution to the $\mu$ problem while significantly reducing the tuning of the ``Little Hierarchy." This results because of a compressed sparticle spectrum ($350 {\rm GeV}$ squarks), a small $\mu$ term, and a Higgs mass lifted above the LEP bound due to a large quartic coupling in the theory. 
  
\subsection{The Two-Singlet Model and mass spectrum}  
  
We will now write the Superpotential that describes the central model of the paper. Let us extend the NMSSM in Gauge Mediation by one additional Singlet ``$S$" having the following interactions
\beqa \label{themodel}
W &=& (\lambda_1 X_1+\lambda_2 X_2)\tilde{\phi_1}\phi_1 +(\kappa_1 Y_1+\kappa_2 Y_2)\tilde{\phi_2}\phi_2 \nonumber \\ &+& \eta_n N\tilde{\phi_2}\phi_1 + \eta_s S \tilde{\phi_1}\phi_2
+\lambda NH_DH_U -\frac{\kappa}{3}S^3.
\eeqa
The first line generates a N-MGM spectrum, and the second line will be responsible for generating the proper interactions that allow Electroweak symmmetry breaking. Since there are Singlets present, we must explain why the superpotential has this form and other operators, not forbidden by gauge symmetries, have not been written down.  An $R$-symmetry is sufficient to do this. If we make the $R$-charge assignments as follows: $R[S] = \frac{2}{3},~R[X_i]=2,~R[Y_i]=r_y,~R[\phi_1]=x,~R[\phi_2]=y$, where $r_y$, $x$, and $y$ are free to be anything. With these choices, the $R$-charges of the other fields are fixed:$ ~R[\tilde{\phi}_1]=2-x,~R[\tilde{\phi}_2]=2-y-r_y$, and $R[N] = \frac{4}{3}+r_y$. Then for $r_y \geq 1$, no renormalizable couplings beyond those that appear in \eq{themodel} are allowed.  In later sections we will assume $F_{Y_1}=F_{Y_2} = 0$, but we will leave them non-zero at the moment for complete generality. 

In this theory we can choose $\Lambda^{(\lambda,\kappa)}_q < \Lambda^{(\lambda,\kappa)}_{\ell}$ such that $m_{sq} \sim 350 ~{\rm GeV}$ and $m_{sl} \sim 150~{\rm GeV}$. Here 
\beq
 \eta_n N \tilde{\phi_2}\phi_1 \equiv \eta^{(T)}_n N \tilde{\phi_2}^{(T)}\phi^{(T)}_1+ \eta^{(D)}_n N \tilde{\phi_2}^{(D)}\phi^{(D)}_1,
 \eeq
 and similarly for the $S$ interaction. 
 We will consider two models. In the first we will consider $\eta^{(T)}_n(M_{GUT})=\eta^{(D)}_n(M_{GUT})$ and parameterize the model in terms of $\eta^{(D)}_n(m_Z)$ where $\eta^{(T)}_n(m_Z)$ can be inferred from GUT relations. In the second case we will consider $\eta^{(T)}_n(M_{GUT})=0$, $\eta^{(D)}_n(M_{GUT}) \not= 0$. 
 Due to the direct Singlet-Messenger interactions, the leading soft masses and interactions for this theory are
\beq \label{soft}
V^{(soft)}=m_N^2|N|^2+m_S^2|S|^2+[bSN-\lambda A_{\lambda}NH_DH_U -\frac{\kappa}{3}A_{\kappa}S^3 + \rm{h.c.}],
\eeq
plus a one-loop supersymmetric mass term for the two Singlets
\beq \label{softmu}
W'= \tilde{\mu}SN.
\eeq
Above the Messenger scale, the model has 9 parameters: 
\beq \label{UVpara}
\left( \lambda, ~\kappa,~ \eta^{(D)}_s,~ \eta^{(D)}_n,~ \Lambda^{(\lambda)}_q, ~\Lambda^{(\kappa)}_q, ~ \Lambda^{(\lambda)}_{\ell}, ~\Lambda^{(\kappa)}_{\ell}, x=\frac{m_{\phi_1}}{m_{\phi_2}} \right).
\eeq
An explicit calculation relates the 6 soft parameters in \eq{soft} and \eq{softmu}, 
\beq
\left( m_N^2,~ m_S^2, ~ b, ~A_{\lambda}, ~A_{\kappa},~ \tilde{\mu} \right), 
\eeq
 to the parameters of \eq{UVpara} as 
\beq \label{soft1}
m^2_N=m_S^2 \left( \frac{\eta_n}{\eta_s} \right)^2= \frac{5\eta^{2}_n}{(16\pi^2)}(\Lambda_1 -\Lambda_2)^2\frac{x^2}{(1-x^2)^3}\left( (1+x^2)\log{x^2}+2(1-x^2) \right),
\eeq
\beq \label{soft2}
b=\frac{5\eta_s\eta_n}{(16\pi^2)}\left[ \frac{x}{(1-x^2)^3}\left( x^4-1-2x^2\log{x^2}\right) (\Lambda_1-\Lambda_2)^2+\frac{x\log{x^2}}{(1-x^2)}\Lambda_1\Lambda_2\right],
\eeq
\beq
\tilde{\mu}=\frac{-5\eta_n\eta_s}{(16\pi^2)}\left( \frac{x}{1-x^2}(\Lambda_2-\Lambda_1)+(x^2\Lambda_2-\Lambda_1)\frac{x\log{x^2}}{(1-x^2)^2}\right),
\eeq
\beq
-\left( \frac{\eta_n}{\eta_s} \right)^2\frac{A_{\kappa}}{3}=A_{\lambda} =\frac{5\eta^{2}_n}{(16\pi^2)}\left( \frac{x}{(1-x^2)}(\Lambda_2-\Lambda_1)+(x^2\Lambda_2-\Lambda_1)\frac{x\log{x^2}}{(x^2-1)^2} \right).
\eeq
Letting $a,b = s,n$, we have
\beqa \label{softfinal}
(\eta_a\eta_b)\Lambda_1\Lambda_2 & =& \frac{3(\eta_a^{(T)}\eta_b^{(T)})}{5} \Lambda_q^{(\kappa)} \Lambda_q^{(\lambda)} +\frac{2(\eta_a^{(D)}\eta_b^{(D)})}{5} \Lambda_{\ell}^{(\kappa)} \Lambda_{\ell}^{(\lambda)}, \\
(\eta_a\eta_b)(\Lambda_1-x^2 \Lambda_2)& = &\frac{3(\eta^{(T)}_a\eta^{(T)}_b)}{5}(\Lambda^{(\lambda)}_{q} - x^2 \Lambda^{(\kappa)}_{q})+ \frac{2(\eta_a^{(D)}\eta_b^{(D)})}{5}(\Lambda^{(\lambda)}_{\ell} -x^2 \Lambda^{(\kappa)}_{\ell}), \\
(\eta_a\eta_b)(\Lambda_1- \Lambda_2) &=& \frac{3(\eta_a^{(T)}\eta_b^{(T)})}{5}(\Lambda^{(\lambda)}_{q} -  \Lambda^{(\kappa)}_{q})+ \frac{2(\eta_a^{(D)}\eta_b^{(D)})}{5}(\Lambda^{(\lambda)}_{\ell} - \Lambda^{(\kappa)}_{\ell}), \\
(\eta_a\eta_b)(\Lambda_1- \Lambda_2)^2 &=& \frac{3(\eta_a^{(T)}\eta_b^{(T)})}{5}(\Lambda^{(\lambda)}_{q} -  \Lambda^{(\kappa)}_{q})^2+ \frac{2(\eta_a^{(D)}\eta_b^{(D)})}{5}(\Lambda^{(\lambda)}_{\ell} - \Lambda^{(\kappa)}_{\ell})^2.
\eeqa
Below the Messenger scale, SUSY is broken and the Electroweak breaking vacuum is determined by the VEVs of the neutral components of the fields. We assume that there is no explicit CP violation in the UV parameters of the model. We will take the 9 potentially complex UV parameters of the model in \eq{UVpara} to be Real. In general this will not be true, but since CP violation in nature is small, we will consider this a reasonable assumption for the scope of this paper. Even without explicit CP violation, spontaneous CP violating effects can arise if physical phases arise in the neutral field VEVs. However, we will always consider vacua where this does not occur.  Assuming CP is conserved in the vacuum, we may work in a basis where the field VEVs are Real and positive and the vacuum potential energy is 
 \beqa
&&V^{(vac)}(s,n,v_u,v_d) = m_N^2n^2+m_S^2s^2+m_{H_U}^2v_u^2 + m_{H_D}^2v_d^2 \nonumber \\  &+& \lambda^2(n^2v_u^2+n^2v_d^2+v_u^2v_d^2) + \kappa^2 s^4 + \frac{g^2+g'^2}{4}(v_d^2-v_u^2)^2 \nonumber \\ &-&2\lambda A_{\lambda}v_dv_un -2\kappa A_{\kappa}s^3+\tilde{\mu}^2(s^2+n^2)+2bsn-2\kappa \tilde{\mu} xs^2+2\lambda \tilde{\mu} s v_d v_u,
\eeqa
where $s = \langle S \rangle$ and $n = \langle N \rangle$. 
Minimizing this potential energy yields the following four minimization conditions. The first two are familiar and relate the Higgs soft masses to $v^2=v_u^2+v_d^2=(174~{\rm GeV})^2$ and $\tan{\beta}=\frac{v_u}{v_d}$:
\beq \label{min1}
\mu^2=\frac{m_{H_D}^2-m_{H_U}^2\tan{\beta}^2}{\tan^2{\beta}-1}-\frac{m_Z^2}{2},
\eeq
\beq\label{min2}
\sin{2\beta}=\frac{2\bar{B}}{m_{H_U}^2+m_{H_D}^2+2\mu^2},
\eeq
with 
\beq
\bar{B}=\lambda A_{\lambda}-\lambda\tilde{\mu}s-\frac{\lambda^2v^2}{2}\sin{2\beta}, ~~~~ \mu= \lambda n.
\eeq
The second two equations relate the Singlet soft mass parameters to the VEVs of $S$ and $N$:
\beq \label{min3}
m_N^2=-\lambda^2v^2+\lambda A_{\lambda}\frac{v^2\sin{2\beta}}{2n}-\tilde{\mu}^2-b\frac{s}{n}+\kappa\tilde{\mu}\frac{s^2}{n},
\eeq
\beq \label{min4}
m_S^2=-2\kappa^2s^2+\kappa A_{\kappa}s-\tilde{\mu}^2-b\frac{n}{s}+2\kappa\tilde{\mu}n-\lambda\tilde{\mu}\frac{v^2\sin{2\beta}}{2s}.
\eeq 
There are 4 neutral CP-even Higgs states and 3 neutral CP-odd Higgs states (the 4th CP-odd state is eaten by the $Z$-boson). It is straightforward to write the $4 \times 4$ CP-even and $3 \times 3$ CP-odd Higgs mass matrices:
\beq \label{cpemass}
M^2_{CPE}=\left( \begin{array}{cccc} m_{h^0}^2& -v^2(\bar{g}-\lambda^2)\cos{4\beta}& \lambda v (2\lambda n-A_{\lambda} \sin{2\beta}) &\lambda v \tilde{\mu} \sin{2\beta}\\
-v^2(\bar{g}-\lambda^2)\cos{4\beta}&m_{H^0}^2 & -\lambda A_{\lambda}v \cos{2\beta}&\lambda v \tilde{\mu} \cos{2\beta}  \\
\lambda v (2\lambda n -A_{\lambda} \sin{2\beta})  &   -\lambda A_{\lambda}v \cos{2\beta}& m_n^2& b-2\kappa \tilde{\mu}s \\
\lambda v \tilde{\mu} \sin{2\beta}&\lambda v \tilde{\mu} \cos{2\beta} &b-2\kappa \tilde{\mu}s & m_s^2
\end{array}
\right),\,
\eeq
with
\beqa \label{mii}
m_{h^0}^2 &=&\bar{g}v^2\cos^2{2\beta}+\lambda^2v^2\sin^2{2\beta}+ \delta m_h^2, \\
m_{H^0}^2 &=& (\bar{g}-\lambda^2)v^2\sin^2{2\beta} + \frac{2\lambda(A_{\lambda}n-s \tilde{\mu})}{\sin{2\beta}},\\
m_n^2 &=& -b\frac{s}{n}+\frac{\lambda A_{\lambda}\sin{2\beta}}{2n} + \frac{\kappa \tilde{\mu}s^2}{n} -\tilde{\mu}^2,\\
m_s^2&=&4\kappa^2s^2 - \kappa A_{\kappa} s- \tilde{\mu}^2 - b\frac{n}{s} -\frac{\lambda \tilde{\mu} \sin{2\beta}}{2s},\\
\bar{g} &=&\frac{(g^2+g'^2)}{2},~~~~~
\delta m_h^2  = \frac{3m_t^2}{4\pi^2v^2}\log{\left( \frac{\bar{m}_{\tilde{t}}^2}{m_t^2} \right)}, 
\eeqa
and
\beqa \label{cpomass}
M^2_{CPO}=\left( \begin{array}{ccc} \frac{2\lambda (A_{\lambda}n -s\tilde{\mu})}{\sin{2\beta}}&\lambda A_{\lambda}v& \lambda v \tilde{\mu} \\
\lambda A_{\lambda}v&-b\frac{s}{n}+\lambda A_{\lambda}\sin{2\beta}\frac{v^2}{2n} +\kappa \tilde{\mu}\frac{s^2}{n}-\tilde{\mu}^2& -b-2\kappa s \tilde{\mu} \\
\lambda v \tilde{\mu}   & -b-2\kappa s \tilde{\mu} & -b\frac{n}{s}+3\kappa A_{\kappa}s-\tilde{\mu}(\tilde{\mu}+\frac{\lambda \sin{2\beta}v^2}{2s}-4n\kappa)\\
\end{array}
\right).  \nonumber \\
\eeqa
Here we have written the mass matrices in the ``Higgs" basis $\left( h^0, H^0, N, S\right)$. Where
\beqa
\left( \begin{array}{c} h^0 \\ H^0 \end{array} \right) = \left( \begin{array}{cc} \cos{\beta} & \sin{\beta} \\
-\sin{\beta} & \cos{\beta} \end{array} \right) 
\left( \begin{array}{c} H^0_D \\ H^0_U \end{array} \right).
\eeqa
Here, the $h^0$ state is the only one coupling to the $Z$.  
Recall that the superpotential operator $\lambda N H_D H_U$ could have two effects. The first is to generate the $\mu$ term if $N$ gets a VEV. From the CP-even mass matrix \eq{cpemass}, we can see that if we neglect $\tilde{\mu}$ and $A_{\kappa}$, the general structure of \eq{simplemodel} emerges in the $N$/$S$ sector when $\frac{s}{x} = \tan{\beta'} \gg 1$, $\tan{\beta}\sim 1$.  This has the effect of giving the $N$ state a small VEV relative to its mass. The VEV generates a $\mu$ term. It is true that $S$ and $N$ will generally get VEVs; for instance, \eq{soft1} and \eq{soft2} imply that $b^2 > m_S^2m_N^2$ is always true when $F_{Y_i} = 0$, which will be true for the parameter points we will consider next. The second effect of $\lambda N H_D H_U$ is to lift the Higgs mass through an increased quartic coupling. Here, this is clearly visible as the $\lambda^2 v^2 \sin^2{2\beta}$ part of the $(M^2_{CPE})_{11}$ entry. The lightest eigenvalue will be strictly less than $m_{h^0}^2$ due to mixing effects, but this mixing effect will be minimized when the mass of the propagating $N$ state is large, which is generally true from our discussion in Section IV. 

In addition to the scalar Higgs spectrum, there are also the associated fermions. The two new Singlets mix with the Higgsinos and gauginos enlarging the Neutralino Mass matrix to a $6 \times 6$ mass matrix. This is simply derived from
\beq
W= \lambda N H_U H_D -\frac{\kappa}{3}S^3 + \tilde{\mu} SN
\eeq
and
\beq
\mathcal{L}_{mass} = -\frac{1}{2} \sum_{ij} \frac{\partial^2W}{\partial \phi_i \partial \phi_j}\psi_i \psi_j + {\rm h.c.}
 \rightarrow -\mathcal{L}_{mass} = \frac{1}{2}\vec{\Psi}^T \mathcal{M}\vec{\Psi} + {\rm h.c.}
\eeq
Writing the matrix in the basis
\beq
\vec{\Psi} = \left( \lambda', ~\lambda^3,~ \psi_{H_U}^0, ~\psi_{H_D}^0,~\psi_N, ~\psi_S \right),
\eeq
we have
\beqa
\mathcal{M} =\left( \begin{array}{cccccc} M_1 & 0 & -m_Z s_{\beta}s_W & m_Zc_{\beta}s_W & 0 & 0 \\
0 & M_2 & m_Zc_Ws_{\beta} & -m_Zc_{\beta}c_W &0&0 \\ 
-m_Z s_{\beta}s_W &  m_Zc_{\beta}s_W & 0 & \mu & \lambda vs_{\beta}& 0 \\
m_Zc_{\beta}s_W & -m_Zc_{\beta}c_W & \mu & 0& \lambda vc_{\beta} & 0 \\
0 & 0 & \lambda vs_{\beta} & \lambda vc_{\beta} & 0 & \tilde{\mu} \\
0 & 0 & 0 & 0 & \tilde{\mu} & -2\kappa \langle S \rangle \end{array} \right) .
\eeqa
The mass eigenbasis ($\chi^0_i$) is given by acting with a rotation matrix $U$ such that
\beq
U \mathcal{M} U^T = \mathcal{M}_D = {\rm diag}(m_{\chi^0_i}), ~~~ U_{ij}\Psi_j = \chi^0_i, ~~~ (U^T)_{ij}\chi^0_j = \Psi_i.
\eeq
 The interaction that arises from the $\lambda N H_D H_U$ term is 
 \beq
 \mathcal{L}_{int} =-\lambda H_U^0 \psi_N \psi_{H_D}  -\lambda H_D^0 \psi_N \psi_{H_U}  + {\rm h.c.}
 \eeq
 In the mass eigenbasis the Lightest CP-even Higgs can be approximated by the light state in the Higgs basis above. Here we neglect the small mixing angle, $\alpha$, mixing that rotates from the states from the Higgs basis to the mass eigenstate basis, $\alpha$ is typically small here because the the heavy CP-even Higgs is heavy. So, 
 \beq 
 \left( \begin{array}{cc} c_{\beta} & -s_{\beta} \\ s_{\beta} & c_{\beta} \end{array} \right) \left( \begin{array}{c} h^0 \\ H^0 \end{array} \right) =  \left(  \begin{array}{c} H_D^0 \\ H_U^0 \end{array} \right). 
 \eeq
 Finally we have 
 \beq
 \mathcal{L}_{int} = y_{\chi^0_{11}} h^0 \chi_1^0\chi^0_1 + {\rm h.c.} ~~~ y_{\chi^0_{11}} = -\lambda [ c_{\beta} [U^T]_{51} [U^T]_{41} + s_{\beta} [U^T]_{51}[U^T]_{31}].
 \eeq
 Typical masses of $\chi^0_1$ are naturally $m_{\chi^0_1}\sim \frac{\lambda^2 v^2}{\mu} \sim 43-53~ {\rm GeV}$. This makes lightest Singlino of the theory kinematically accessible as long as $m_{h^0} > 2 m_{\chi_1^0}$, which will be true for $m_{h^0} > 110~{\rm GeV}$. This Higgs is Standard Model-like when it is produced, but it's decays are altered. We can write the Branching Ratio,
 \beq 
 {\rm BR}(h^0 \rightarrow \bar{b}b) \sim \frac{ \Gamma(h^0 \rightarrow \bar{b}b)}{\Gamma(h^0 \rightarrow \bar{b}b) + \Gamma(h^0 \rightarrow \bar{\chi^0}_1\chi^0_1)},
 \eeq
 where we estimate
 \beq \label{decay}
 \Gamma( h^0 \rightarrow \bar{b}b) \sim y_b^2~~~~  {\rm and} ~~~ ~\Gamma( h^0 \rightarrow \bar{\chi^0}_1\chi^0_1) \sim y_{\chi^0_{11}}^2\left( 1-\frac{4m^2_{\chi_1^0}}{m^2_{h^0}}\right)^{\frac{3}{2}}.
 \eeq
 This decay mode will be important when we consider regions of parameter space where the Higgs mass is $m_{h^0} = 110~ {\rm GeV}$. This is reminiscent of the scenarios investigated in \cite{Chang:2005ht, Chang:2008cw}.

Finally, we mention a feature of the Chargino mass matrix. It takes the form
\beqa
M_{chargino}=\left( \begin{array}{cc} M_{\lambda} & \sqrt{2}m_W \sin{\beta} \\  \sqrt{2}m_W \cos{\beta} &-\mu \end{array} \right).
\eeqa 
In this model the  ``sign of the $\mu$ term is negative", meaning the sign of the diagonal entries of the Chargino mass matrix are opposite while the off diagonal entries are positive. We can say this more precisely in the following way: choosing the sign of the gaugino masses as positive determines the sign of $A_{\lambda}$, ($A_{\lambda} > 0$ in the notation used in this paper). Also, our choice to work in a basis with $\tan{\beta} >0$ determines that $B\mu=A_{\lambda}\lambda \langle N \rangle > 0$.  Since, $\mu = \lambda \langle N \rangle$ then $\mu > 0$. This is an important feature when considering low values of $\mu$ because a natural cancellation exists in the mixing of the gaugino/higgsino states. This aides in keeping the lightest state above the LEP bound of $103~ {\rm GeV}$ if both $\mu$ and $M_2$ are small. 

To summarize, in this section we have derived from the Supersymmetric theory of \eq{themodel} the entire neutral Higgs sector scalar lagrangian, its minimization conditions, and the mass matrices for the physical fields after Supersymmetry breaking has been communicated to the Higgs sector via direct Messenger Singlet interactions.  We have seen that it naturally incorporates a structure in the Singlet sector very similar to that described in Section II. Also, as discussed in Section II, this model would seem to be capable of solving the $\mu/B\mu$ problem and alleviating the fine-tuning associated with Electroweak symmetry breaking.  In the next subsection, we investigate under what assumptions the model of \eq{themodel} can actually accomplish this, by finding viable regions of the model's parameter space.

\subsection{The Model Parameters}

We now choose a set of Benchmark parameters and investigate the mass spectrum of the two-Singlet model. We will see that it realizes a Higgs heavier than $114.4~ {\rm GeV}$ when $\tan{\beta} \sim 1.5$. The two-Singlet model in \eq{themodel} has 9 parameters: \eq{UVpara}. Five of these nine parameters determine squark, gaugino, and slepton masses arising due to the two-Messenger Sector: $\left(  \Lambda^{(\kappa)}_q, \Lambda^{(\kappa)}_{\ell}, \Lambda^{(\lambda)}_q, \Lambda^{(\lambda)}_{\ell}, x \right)$.  In what follows we will set $\Lambda^{(\kappa)}_q = \Lambda^{(\kappa)}_{\ell} = 0$ because it simplifies our analysis and it is not unreasonable to assume that the Messengers $\phi_2$ and $\tilde{\phi}_2$ couple to a Spurion having dynamics different from that of the $X_i$ fields that carry $F$-term VEVs as in \eq{sup2}. For simplicity we set
\beq
W=M_2\tilde{\phi}_2\phi_2, ~~ {\rm with} ~~ M_2 = \kappa_1\langle Y_1 \rangle + \kappa_2 \langle Y_2 \rangle.
\eeq
The parameter ``$x=\frac{M_1}{M_2}$" only affects the values of the soft parameters by an order one amount. So the phenomenology is weakly dependent on this parameter except in the extreme limits that $x\rightarrow 0$ or $x \rightarrow \infty$. For this reason we will take $x = 0.5$ in numerical results.  

Of the remaining six parameters, $\Lambda^{(\lambda)}_q$ and $\Lambda^{(\lambda)}_{\ell}$ determine the squark and slepton masses. The main interest of this paper is to investigate the possibility that the Higgs mass is lifted by the dynamics of a tree-level modification to the Higgs quartic coupling rather than the usual radiative stop contribution. For this reason we will only consider $m_{sq} \sim 350~{\rm GeV}$ and $m_{sl} \sim 150~{\rm GeV}$, these constraints determine $\Lambda^{(\lambda)}_q \sim 15-18~{\rm TeV} < \Lambda^{(\lambda)}_{\ell} \sim 60-120~{\rm TeV}$. 

Fixing the 5 parameters: $\left(  \Lambda^{(\kappa)}_q, \Lambda^{(\kappa)}_{\ell}, \Lambda^{(\lambda)}_q, \Lambda^{(\lambda)}_{\ell}, x \right)$, the Electroweak minimization conditions, \eq{min1}-\eq{min4}, determine one remaining parameter of the model.  We choose this to be the parameter $\kappa$. This essentially exchanges the value of $m_Z$ for $\kappa$. At this point we have reduced the number of free parameters from nine to three: $\left( \eta^{(D)}_n, \eta^{(D)}_s, \lambda \right)$. 

Since the theory does not have heavy stops, we must rely on a large $\lambda$ to lift the Higgs mass. In this paper we will only consider theories perturbative to the GUT scale. A large $\lambda$ can be realized in a scenario where the gauge couplings are as large as possible without hitting Landau poles at the GUT scale. In the two models we will consider, the model with GUT relations will have $N=2$ Messengers and the model without GUT relations will have $N=4$ Messengers. Extra chiral matter helps keep $\lambda$ perturbative to the GUT scale.  We find that the size of $\lambda$ is also very sensitive to the top Yukawa coupling and therefore on $\tan{\beta}$. Essentially, as $\tan{\beta}$ decreases in value, the top Yukawa increases, renormalizing $\lambda$ to larger values in the UV. We will require perturbativity to the GUT scale by requiring $ \frac{\lambda(M_{GUT})}{4\pi}< 0.3$. We will impose this restriction by fixing  $\lambda(m_Z) = 0.65$ and $\lambda(m_Z) = 0.7$ in the models with GUT relations and without GUT relations, respectively, and if a vacuum solution arises in which $\frac{\lambda(M_{GUT})}{4\pi} > 0.3$, we will not consider this solution viable. Renormalization Group Equations for $\lambda$, the gauge couplings, and the Singlet-Messenger Yukawa couplings can be easily inferred from the appendix of \cite{Delgado:2007rz}. 

Finally, after fixing $\lambda(m_Z)$, we have two free parameters: $\left( \eta^{(D)}_s, \eta^{(D)}_n \right)$.  Recall from our discussion in Section II that the structure giving the field $N$ a small VEV relative to its mass relies on $\tan{\beta'} \gg 1$ in \eq{betaprime}. This can arise if $\frac{\eta^{(D)}_s}{\eta^{(D)}_n} \ll 1$.  This requires the Yukawa coupling of $\eta^{(D)}_s$ to be small. We would like to emphasize that the smaller Yukawa does not have to be hierarchically small, just somewhat small, and so in what follows we will set $\epsilon=\frac{\eta^{(D)}_s}{\eta^{(D)}_n} =0.1$. 

Now we have reduced the nine parameter model to one parameter which we will take to be $\eta^{(D)}_n$. In what follows we will simplify notation and set $\eta^{(D)}_n=\eta_n$. we can now evaluate the phenomenological parameters $\mu$, $m_{Higgs}$ (the lightest CP-even Higgs mass eigenvalue), and $\tan{\beta}$ in terms of $\eta_n$ for each of the two models. 

\subsection{GUT relations and $N$=2 Messengers} 

We now consider a Two-Singlet model with $N=2$ Messengers and GUT relations between the doublet and triplet couplings. Reasoning as above, our Benchmark Point for the model is
\beq \label{parame}
 \Lambda^{(\lambda)}_{\ell} = 85~{\rm TeV},  ~~\Lambda^{(\lambda)}_{q} = 17.5 {\rm TeV}, ~~\frac{\eta_s}{\eta_n} = 0.1, ~~ \lambda = 0.65, ~~ x= 0.5, ~~\Lambda^{(\kappa)}_{q} =\Lambda^{(\kappa)}_{\ell} = 0.
\eeq

We also take the Messenger mass scale to be $M= 3 \Lambda_{\ell}$ corresponding to Low-Scale Gauge Mediation. The phenomenological parameters  $m_{Higgs}$, $\tan{\beta}$, and $\mu$ are shown in Fig. 1 , Fig. 2, and  Fig. 3 as a function of $\eta_n$, the doublet-Singlet Yukawa coupling. $\kappa$ is an output for each $\eta_n$ and on average is close to $\kappa \sim -0.15$ for all solutions. The vacuum ceases to be globally stable when $\kappa > -0.02$. $\kappa$ does not achieve those dangerous values for any of our parameter points. 

\begin{figure} 
\label{hmass}
\includegraphics{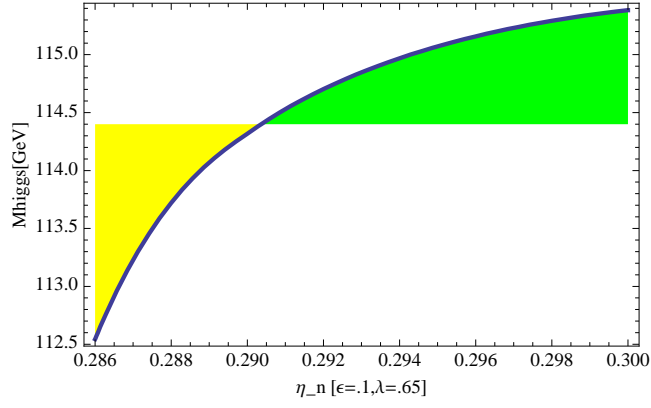} 
\caption{The Green Region (dark shaded) shows the allowed range of $\eta_n$ for which $m_{Higgs}$ is above $114.4~{\rm GeV}$:  $\eta_n > .290$.}
\end{figure}

\begin{figure} 
\label{tb}
\includegraphics{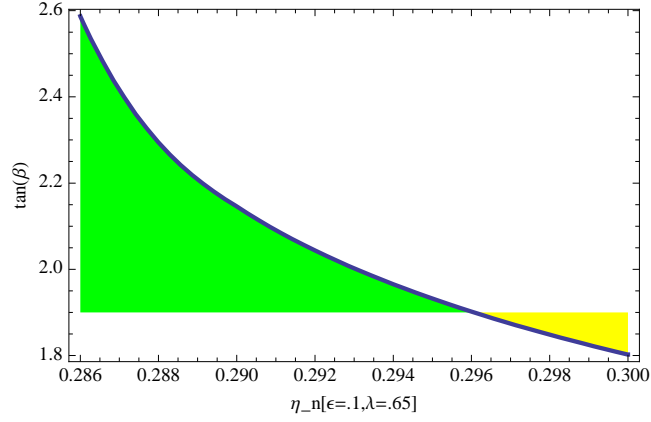} 
\caption{The Green Region (dark shaded) shows the allowed range of $\eta_n(m_Z)$ for which $\tan{\beta}< 1.9$, $\eta_n< .296$, which corresponds to the region of parameter space where $\frac{\lambda_{GUT}}{4\pi}< 0.3$. When $\eta_n(m_Z)=0.29$, $\eta_n(m_{GUT})= 0.57$.}
\end{figure}

\begin{figure} 
\label{mu}
\includegraphics{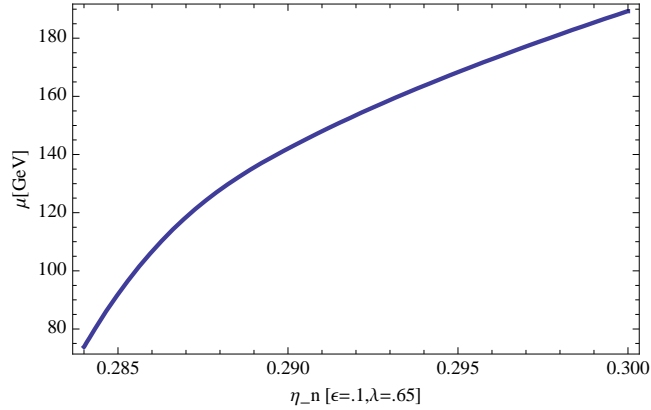} 
\caption{In the allowed regions of parameter space $\mu= [140 {\rm GeV}, 170 {\rm GeV}]$.}
\end{figure}

The parameters in \eq{parame} are input at the Messenger scale.  \Eq{gluinoformula} and \eq{squarkformula} determine the soft masses at the Messenger scale, and \eq{soft1} - \eq{softfinal}  determine the soft SUSY breaking parameters in \eq{soft} at the Messenger scale. In general these parameters should run via RG evolution from the Messenger scale to the Electroweak scale where the minimization conditions and particle masses should be evaluated. However, since we are taking the Messenger Scale to be low, RG running from the Messenger to the Electroweak scale changes the soft parameters by a small amount.  There are however, two effects of RG running that are important despite the low Messenger scale, due to the two largest Yukawa couplings of the theory. The first and most important is due to the top yuakwa coupling. As discussed in Section I, it significantly renormalizes the up-type Higgs soft mass parameter  ($m^2_{H_U}$), as well as contributed radiatively to the physical mass of the SM-like CP-even Higgs. The second important radiative effect is the new large coupling in the theory, $\lambda$. This somewhat large Yukawa coupling renormalizes both soft Higgs masses by an amount that is similar in magnitude but opposite in sign to the finite two-loop threshold that arises from integrating out the Messengers,
\beq
\delta m^2_{H_U}=\delta m^2_{H_D}= -\frac{\lambda^2}{8\pi^2}m_N^2\log{\frac{M}{m_N}},
\eeq
where $M$ is the Messenger scale and $m_n$ is the mass of the heavy propagating Field in $N$. For the models we consider it is typical that $m_n \sim 1.5~{\rm TeV}$. It is important to ask if this exacerbates the ``Little Hierarchy" problem because of this large Renormalization effect. It is very similar in form and magnitude to the renormalization in MGM from a heavy stop. But this renormalization is different in two ways. First, $\lambda < y_t$ and N is not colored, therefore the soft mass renormalization is roughly a factor of 6 less severe than it would be for $1.5~ {\rm TeV}$ stops.  Secondly, the stop mass renormalization renormalizes only $m^2_{H_U}$, whereas a large $\lambda$ renormalizes both Higgs soft masses by an equal amount. The combination of these two facts are why this renormalization does not spoil this model as a solution to the ``Little Hierarchy."  

After these RG effects are included as corrections to the tree-level mass matrices and soft parameters for the model, we evaluate the four minimization conditions  \eq{min1}-\eq{min4}. These four conditions determine $\kappa$, $\tan{\beta}$, $\langle N \rangle$, and $\langle S \rangle$. These parameters determine the lightest CP-even Higgs mass  via \eq{cpemass} as well as $\mu$.  

The results in Figure 1 are easy to interpret. As $\eta_n$ increases, so does $A_{\lambda}$. This soft parameter is essentially the $B\mu$-term of the typical MSSM Higgs potential. A large $A_{\lambda}$ induces a smaller $\tan{\beta}$ and shifts the main contribution to the lightest CP-even Higgs mass from $m_Z\cos{2\beta}$ to $\lambda v \sin{2\beta}$, thereby allowing a large tree-level Higgs mass.

The allowed region of values for $\eta_n$ are the regions of Fig. 1 where $m_{Higgs} > 114.4 ~{\rm GeV}$ and Fig. 2 where $\tan{\beta} > 1.9$. This is the overlap of the green (dark shaded) regions of Fig. 1 and Fig. 2. If $\tan{\beta} < 1.9$ the large top Yukawa determines $\frac{\lambda_{GUT}}{4\pi}>0.3$ and we do not consider these solutions. A non-zero but small region of allowed $\eta_n$ exists for $\eta_n = [.290,.296]$. The value of $\mu$ in this allowed regions is as low as $\mu = [140 ~{\rm GeV} , 170 ~ {\rm GeV}]$. 

Applying the rough tuning measure $T\sim \frac{\mu^2}{\frac{m_Z^2}{2}} \sim 4.8-7.1$ this corresponds to roughly a $14\%-20\%$ tuning to get the correct Electroweak scale. We consider this a very minimal amount of tuning of the Electroweak Minimization conditions. However, clearly for this model there is a roughly $1\%$ tuning to find $\eta_n$ in the region where $m_{Higgs}> 114.4~{\rm GeV}$. For this particular model the tuning has shifted from tuned minimization conditions to a tuned parameter space for a phenomenologically allowed value of $\eta_n$. The region of allowed $\eta_n$ remains small or disappears as one varies the other parameters in \eq{parame}. For instance, if $\Lambda_q^{\lambda} > 18.5~{\rm TeV}$ or $\lambda < 0.6 $ the allowed region of parameter space with with $m_{Higgs} > 114.4 ~{\rm GeV}$ disappears.   

However, while any model with $m_{Higgs} > 114.4 ~ {\rm GeV} $ is safe from the LEP bound, the LEP bound does not apply to a Higgs with non-standard model-like production or decay. In the two-Singlet Model, Higgs Boson production proceeds as in the Standard Model, but for our Benchmark parameters,  one finds that applying \eq{decay} to the vacuum solutions where $m_{Higgs} >110 ~{\rm GeV}$ gives $BR(h^0 \rightarrow \bar{b}b) < 0.1$ due to the presence of the additional Singlino decay channel. This suppressed decay channel appears to persist as we deviate from our Benchmark parameter point, but we have not yet systematically checked the value of $BR(h^0 \rightarrow \bar{b}b)$ in all parameter points in Fig. 4. Regardless, it is clear that when analyzing the parameter space of this model, some regions of parameter space will likely allow $ 114.4 ~{\rm GeV} > m_{Higgs} > 110 ~{\rm GeV}$. This will enlarge the allowed parameter space and so we also plot this range of Higgs masses in Fig 4. There we plot the allowed region of $\eta_n$ as a function of $\Lambda_{\ell}$ and include in green (light shading) the region where $110~{\rm GeV} <m_{Higgs} <114.4~{\rm GeV}$. From Figure 4, we see that even with a lighter Higgs, the allowed region of parameter space is still somewhat small ($ 3\%$).  A more serious study of this Higgs decay mode is required to determine whether the $110 {\rm GeV} - 114.4 {\rm GeV} $ window is allowed or excluded by current experiments, and this will be left for future analysis. 

\begin{figure} 
\label{mu}
\includegraphics{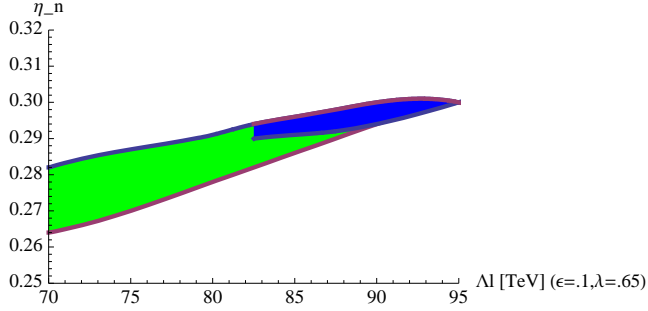} 
\caption{The region of allowed $\eta_n$ as a function of $\Lambda_{\ell}$. Here the other parameters are the same as in the Benchmark point. The Blue (dark shaded) region is where $m_{Higgs} > 114.4~{\rm GeV}$ and the Green (light shaded) region is where $ 114.4 ~{\rm GeV} > m_{Higgs} > 110 ~{\rm GeV}$. As $\Lambda_{\ell} = [70~{\rm TeV}, 95~{\rm TeV}]$, $m_{sq} =  [360~{\rm GeV}, 400~{\rm GeV}]$. }
\end{figure}

Despite the smallness of the allowed parameter space, it is the major source of tuning in the theory, and it is the same as typical tunings found in models of Minimal Gauge Mediation. For this reason we consider this model just as theoretically compelling as Minimal Gauge Mediation and phenomenologically more compelling due to the presence of the light sparticle states.

\subsection{No GUT relations and $N=4$ Messengers} 

If we are free to treat the doublet and triplet Messenger couplings independently, as if they originate from \emph{different} GUT multiplets, the two-Singlet model becomes significantly less constrained. In the previous section we discovered that the allowed parameter space for the Model with GUT relations is quite small. This was due to two effects, both of which are related to Renormalization. The first is that in the previous model, there were only $N=2$ flavors of Messengers. Adding chiral matter to the theory causes the gauge couplings to become larger at the GUT scale, this slows the RG running of  $\lambda$ towards a Landau Pole in the UV and permits a large value for $\lambda(m_Z)$ while maintaining $\frac{\lambda(M_{GUT})}{4\pi}< 0.3$. For $N=2$ Messengers the contribution of the new chiral matter to the gauge coupling RGE is still quite small. The second effect reducing the parameter space is that the GUT relations imply that $\eta_n^{(T)}(m_Z) > \eta_n^{(D)}(m_Z)$. This is due to the tendency of the triplet to have a more negative beta function than that of the doublet. The presence of a large triplet coupling makes the beta function for $\lambda$ more positive and therefore makes a large value for $\lambda(m_Z)$ more difficult to achieve. Also, since we are also working with $\Lambda_q < \Lambda_{\ell}$ we find that it is difficult to make the $A_{\lambda}$ term, and thus the $B\mu$ term, large enough to get a small $\tan{\beta}$ when there are GUT relations between couplings.  This reduces the parameter space where the Higgs is heavier that $114.4 ~{\rm GeV}$. 

We can improve these problems with the following scenario. Let us assume that there are two vector-like pairs of Messengers as in the previous model, but let the doublet and triplet couplings be completely independent. In principle this could originate from a mechanism similar to the mechanism that is responsible for making the Higgs triplets heavy. Now we can turn off all of the triplet-Singlet couplings, setting $\eta_n^{(T)}(m_Z)=\eta_s^{(T)}(m_Z)=0$. Since $\eta_n^{(D)}(m_Z)$ contributes less to the positivity of $\lambda$'s beta function and generates $A_{\lambda}$ via  $\Lambda_{\ell}$, $\lambda(m_Z)$ can be larger in such models. We also add two additional vector-like pairs of Messengers with a supersymmetric mass in order to increase the negativity of the beta function of $\lambda$ from gauge interactions. This could also be accomplished by simply putting a flavor index on the two sets of vector-like Messengers. Since our notation up until this point has only assumed one set of Messengers as the communicators of SUSY breaking to the MSSM, we will consider the case where the extra Messengers are merely spectators, not participating in any interactions other than gauge interactions. We expect the two situations give approximately equivalent results. Strictly speaking, our results will only apply to the case where the extra two vector-like chiral fields are spectators, having a supersymmetric mass like $\tilde{\phi}_2 \phi_2$. The Messenger number is: $N = 4$. We emphasize that the extra two Messengers serve no purpose other than contributing to the renormalization of $\lambda$.

In this model, our Benchmark parameters are:

\beq \label{parame2}
 \Lambda^{(\lambda)}_{\ell} = 80~{\rm TeV},  ~~\Lambda^{(\lambda)}_{q} = 15 ~{\rm TeV}, ~~\frac{\eta^{(D)}_s}{\eta^{(D)}_n} = 0.1, ~~ \lambda = 0.7, ~~ x= 0.5, ~~\Lambda^{(\kappa)}_{q} =\Lambda^{(\kappa)}_{\ell} = 0~~~ \eta_n^{(T)} = 0
\eeq

For this set of parameters the values of the lightest CP-even Higgs mass , $\tan{\beta}$, and $\mu$ are shown in Figure 5, Figure 6, and Figure 7.  

\begin{figure} 
\includegraphics{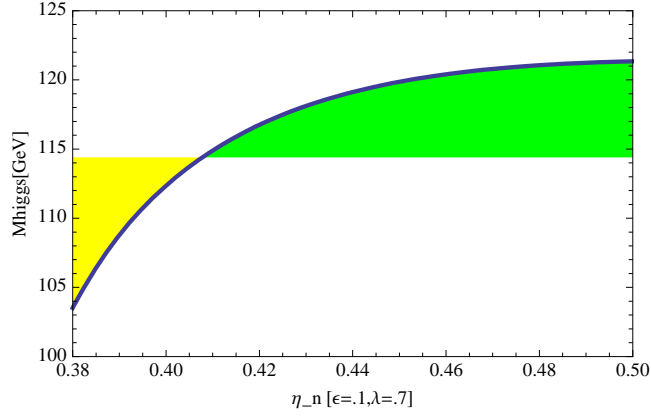} 
\caption{$m_{Higgs}$ as a function of $\eta_n$. The Green (dark shaded) region indicates the range of $\eta_n$ where $m_{Higgs} > 114.4~{\rm GeV}$: $\eta_n >0 .41$. }
\end{figure}

\begin{figure} 
\includegraphics{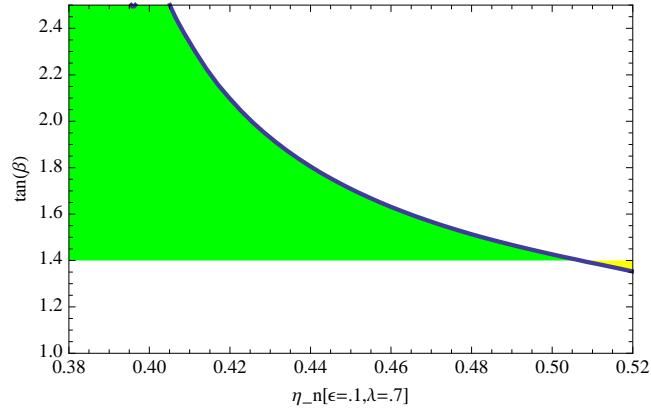} 
\caption{$\tan{\beta}$ as a function of $\eta_n$. The Green (dark shaded) region indicates the region of $\eta_n$ where $\tan{\beta} < 1.9$ which corresponds to $\frac{\lambda_{GUT}}{4\pi} < 0.3$: $ \eta_n < 0.51$.}
\end{figure}

\begin{figure} 
\includegraphics{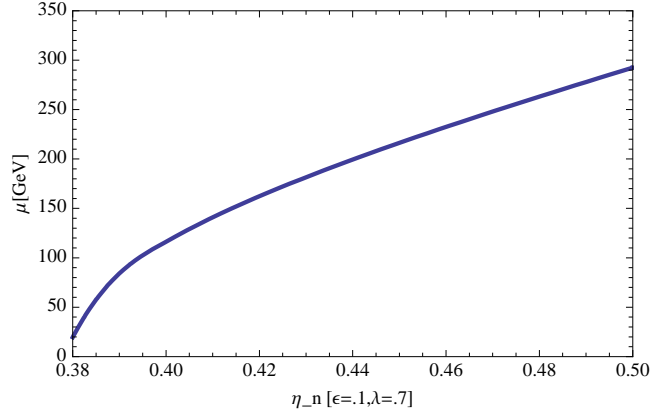} 
\caption{$\mu$ as a function of $\eta_n$. $\mu = [125~{\rm GeV}, 325~{\rm GeV}]$ in the region of allowed parameter space.}
\end{figure}

\begin{figure} 
\includegraphics{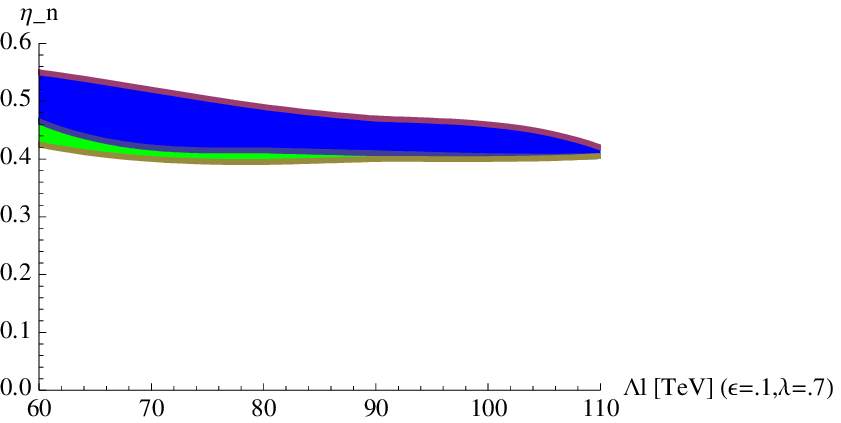} 
\caption{The Blue (dark shaded) region indicates the allowed $\eta_n$ as a function of $\Lambda_{\ell}$ for which $m_{Higgs} > 114.4 ~{\rm GeV}$. The Green (light shaded) region indicates the allowed $\eta_n$ as a function of $\Lambda_{\ell}$ for which $114.4 ~{\rm GeV} > m_{Higgs} > 110~{\rm GeV}$. Here the other parameters are the same as in the Benchmark point. As $\Lambda_{\ell} = [60~{\rm TeV}, 110~{\rm TeV}]$, $m_{sq} =  [320~{\rm GeV}, 370~{\rm GeV}]$. }
\end{figure}

From Figure 5 and Figure 6, we see that this model admits a sizable region of parameter space where $m_{Higgs} > 114.4 ~{\rm GeV}$. In this case $\eta_n = [0.41, 0.51]$.  Figure 7 reveals that this model has a small $\mu$ term: $\mu = [125~{\rm GeV}, 300~{\rm GeV}]$. In order to get a better sense of the parameter space, we plot in Figure 8 the allowed region of $\eta_n$ as a function of $\Lambda_{\ell}$ and include in Green (light shading) the region where $110~{\rm GeV}< m_{Higgs} < 114.4~{\rm GeV}$. An important point is that for $\Lambda_{\ell} > 90~{\rm TeV}$ the lower bound on $\eta_n$ comes from requiring $\mu > 100~{\rm GeV}$ rather than the Higgs mass bounds. For these regions of parameter space the model admits a $\mu$ term of truly minimal value. Here we see that the tuning is roughly a $10\%$ tuning in the parameter space of $\eta_n$.  Applying a simple tuning measure to the Electroweak minimization conditions: $T= \frac{\mu^2}{\frac{m_Z^2}{2}}$, we find that $T\sim 2.5-22$ for $\mu =  [125~{\rm GeV}, 300~{\rm GeV}]$, which corresponds to a tuning in the Electroweak minimization conditions of $40\%-4.5\%$. We conclude that on average there are roughly two  $10\%-20\%$ tunings in this model.  

\section{Conclusions and Outlook}

In this paper we have presented a model of Gauge Mediation that realizes a compressed spectrum (light squarks), a small dynamically generated $\mu$ term, and SM-like Higgs with mass $m_{Higgs}> 114.4 ~{\rm GeV}$. Because the theory has a small $\mu$ term, the Electroweak minimization conditions do not suffer from a ``Little Hierarchy" as in typical Minimal Gauge Mediation. In a model where the Messengers have GUT relations between their couplings, there is only a small region ($1\%$) of parameter space where $m_{Higgs}> 114.4 ~{\rm GeV}$. The region with $m_{Higgs}> 110 ~{\rm GeV}$  is somewhat larger but still only amounts to roughly $(3\%)$ of parameter space. In the model without GUT relations between the Messenger couplings, the region of allowed parameter space with $m_{Higgs}> 114.4 ~{\rm GeV}$ is larger $(10\%)$. The smallness of the $\mu$ term relies on a light stop scenario. Since a compressed sparticle spectrum is not possible in MGM, Non-Minimal Gauge Mediation has been crucial to allowing this phenomenology.  Another important aspect of this theory was mass scale associated with the Singlet soft mass ($m_N^2$). This also relied on the fact that Gauge Mediation was Non-Minimal because in this case an intermediate scale was generated for the Singlet mass ($\mathcal{O}(1-2 {\rm TeV})$). This intermediate mass scale helps the newly generated quartic coupling lift the lightest Higgs mass above $114.4 ~{\rm GeV}$ by suppressing the mixing between the Singlet and the lightest CP-even Higgs state. 

The mass spectrum of these models has some notable features. The theory has a light Gluino, $M_g < 200 ~{\rm GeV}$. Possibly the most interesting feature is that the NLSP is  Singlino, that typically has a mass $m_{\chi_1^0}\sim 50 ~{\rm GeV}$ and can suppress the $h^0 \rightarrow \bar{b}b$ Branching Ratio. If the Higgs decays predominately to the NLSP Singlino, and the Gauge Mediated scale is low, then the final state of the Higgs decay will be two non-pointing photons and missing energy rather than invisible Higgs decays\footnote{ I would like to thank David Morrissey for helping me understand this point}. We do not know of any Higgs study for such a decay mode.  

Since we are working in the context of Low-Scale Gauge Mediation, we have neglected sub-leading Renormalization effects in running from the Messenger Scale to the Electroweak scale, but included the two dominant RG effects from the two large Yukawa couplings in the theory $\lambda$ and $y_{top}$. We have included the stop radiative contribution to the lightest CP-even Higgs mass. Although it is smaller than in MGM, it is an important and leading radiative contribution, being $\mathcal{O}(y^4_{top})$. We have not included radiative contributions to the lightest CP-even mass of sub-leading importance of $\mathcal{O}(y_{top}^2\lambda^2)$ and $\mathcal{O}(\lambda^4)$. The effects of renormalization will change the results of this paper if this two-Singlet Model is used in the context of High Scale Gauge Mediation where the dominant contribution to soft masses comes from RG effects. Our results are only valid for Low-Scale Models.  

We have looked only at the most minimal two parameter extension of Minimal Gauge Mediation, it would be interesting to investigate the Higgs mass bounds as one allows for even more general models of Gauge Mediation. One constraint that we have imposed in this work is perturbativity to the GUT scale. When coupling Singlets to Messengers, it is necessary that the off-diagonal Messenger bilinear combines as a gauge singlet. Models of Gauge Mediation with more than two parameters whose mass spectrum looks distinctly different than the two parameter model likely require Messengers transforming as a ${\bf 10}$ and ${\bf \bar{10}}$ rather than simply more ${\bf 5}$'s and ${\bf \bar{5}}$'s.  In order to have off-diagonal couplings to these Messengers one would require at least two vector-like pairs of them. The minimal effective Messenger number of such a theory is $N=6$. such a theory would have to have high scale Messengers in order to remain perturbative to the GUT scale, and then the effects of RG running would be very important. It unclear whether or not there would be allowed regions of parameter space for such a theory and what the tuning from the ``Little Hierarchy" would be in such a scenario. 

Finally we speculate that it might be possible to think of the two Singlets of this model as forming a vector-like pair under an extra $U(1)$ gauge symmetry. The trilinear coupling of $S$ would not be allowed in such a model, but a quartic coupling would be generated from the D-term potential. It seems plausible that such a model would have many of the features of the two-Singlet model we have investigated here, and for a sufficiently weak gauge coupling for this group, one would expect to maintain a large $\tan{\beta}'$ in the theory. Other subtleties would arise due to the fact that the Higgs and Standard Model particles would have to carry charge under the new broken $U(1)$. Experimental bounds for heavy $Z'$ bosons would apply to such models. We leave the study of the viability of such models for future work.

\section*{Acknowledgements}

I am grateful for a number of illuminating conversations with Michael Dine, David Morrissey, David Poland, and Lisa Randall.  This Research is supported by the National Science Foundation under Grant PHY-0244821 and PHY-0804450 as well as by the Harvard Center for the Fundamental Laws of Nature.


\begin{thebibliography}{45}
\expandafter\ifx\csname natexlab\endcsname\relax\def\natexlab#1{#1}\fi
\expandafter\ifx\csname bibnamefont\endcsname\relax
  \def\bibnamefont#1{#1}\fi
\expandafter\ifx\csname bibfnamefont\endcsname\relax
  \def\bibfnamefont#1{#1}\fi
\expandafter\ifx\csname citenamefont\endcsname\relax
  \def\citenamefont#1{#1}\fi
\expandafter\ifx\csname url\endcsname\relax
  \def\url#1{\texttt{#1}}\fi
\expandafter\ifx\csname urlprefix\endcsname\relax\def\urlprefix{URL }\fi
\providecommand{\bibinfo}[2]{#2}
\providecommand{\eprint}[2][]{\url{#2}}

\bibitem[{\citenamefont{Dine et~al.}(1981)\citenamefont{Dine, Fischler, and
  Srednicki}}]{Dine:1981za}
\bibinfo{author}{\bibfnamefont{M.}~\bibnamefont{Dine}},
  \bibinfo{author}{\bibfnamefont{W.}~\bibnamefont{Fischler}}, \bibnamefont{and}
  \bibinfo{author}{\bibfnamefont{M.}~\bibnamefont{Srednicki}},
  \bibinfo{journal}{Nucl. Phys.} \textbf{\bibinfo{volume}{B189}},
  \bibinfo{pages}{575} (\bibinfo{year}{1981}).

\bibitem[{\citenamefont{Dimopoulos and Raby}(1981)}]{Dimopoulos:1981au}
\bibinfo{author}{\bibfnamefont{S.}~\bibnamefont{Dimopoulos}} \bibnamefont{and}
  \bibinfo{author}{\bibfnamefont{S.}~\bibnamefont{Raby}},
  \bibinfo{journal}{Nucl. Phys.} \textbf{\bibinfo{volume}{B192}},
  \bibinfo{pages}{353} (\bibinfo{year}{1981}).

\bibitem[{\citenamefont{Dine and Fischler}(1982)}]{Dine:1981gu}
\bibinfo{author}{\bibfnamefont{M.}~\bibnamefont{Dine}} \bibnamefont{and}
  \bibinfo{author}{\bibfnamefont{W.}~\bibnamefont{Fischler}},
  \bibinfo{journal}{Phys. Lett.} \textbf{\bibinfo{volume}{B110}},
  \bibinfo{pages}{227} (\bibinfo{year}{1982}).

\bibitem[{\citenamefont{Nappi and Ovrut}(1982)}]{Nappi:1982hm}
\bibinfo{author}{\bibfnamefont{C.~R.} \bibnamefont{Nappi}} \bibnamefont{and}
  \bibinfo{author}{\bibfnamefont{B.~A.} \bibnamefont{Ovrut}},
  \bibinfo{journal}{Phys. Lett.} \textbf{\bibinfo{volume}{B113}},
  \bibinfo{pages}{175} (\bibinfo{year}{1982}).

\bibitem[{\citenamefont{Alvarez-Gaume et~al.}(1982)\citenamefont{Alvarez-Gaume,
  Claudson, and Wise}}]{AlvarezGaume:1981wy}
\bibinfo{author}{\bibfnamefont{L.}~\bibnamefont{Alvarez-Gaume}},
  \bibinfo{author}{\bibfnamefont{M.}~\bibnamefont{Claudson}}, \bibnamefont{and}
  \bibinfo{author}{\bibfnamefont{M.~B.} \bibnamefont{Wise}},
  \bibinfo{journal}{Nucl. Phys.} \textbf{\bibinfo{volume}{B207}},
  \bibinfo{pages}{96} (\bibinfo{year}{1982}).

\bibitem[{\citenamefont{Dimopoulos and Raby}(1983)}]{Dimopoulos:1982gm}
\bibinfo{author}{\bibfnamefont{S.}~\bibnamefont{Dimopoulos}} \bibnamefont{and}
  \bibinfo{author}{\bibfnamefont{S.}~\bibnamefont{Raby}},
  \bibinfo{journal}{Nucl. Phys.} \textbf{\bibinfo{volume}{B219}},
  \bibinfo{pages}{479} (\bibinfo{year}{1983}).

\bibitem[{\citenamefont{Dine and Nelson}(1993)}]{Dine:1993yw}
\bibinfo{author}{\bibfnamefont{M.}~\bibnamefont{Dine}} \bibnamefont{and}
  \bibinfo{author}{\bibfnamefont{A.~E.} \bibnamefont{Nelson}},
  \bibinfo{journal}{Phys. Rev.} \textbf{\bibinfo{volume}{D48}},
  \bibinfo{pages}{1277} (\bibinfo{year}{1993}), \eprint{hep-ph/9303230}.

\bibitem[{\citenamefont{Dine et~al.}(1995)\citenamefont{Dine, Nelson, and
  Shirman}}]{Dine:1994vc}
\bibinfo{author}{\bibfnamefont{M.}~\bibnamefont{Dine}},
  \bibinfo{author}{\bibfnamefont{A.~E.} \bibnamefont{Nelson}},
  \bibnamefont{and} \bibinfo{author}{\bibfnamefont{Y.}~\bibnamefont{Shirman}},
  \bibinfo{journal}{Phys. Rev.} \textbf{\bibinfo{volume}{D51}},
  \bibinfo{pages}{1362} (\bibinfo{year}{1995}), \eprint{hep-ph/9408384}.

\bibitem[{\citenamefont{Dine et~al.}(1996)\citenamefont{Dine, Nelson, Nir, and
  Shirman}}]{Dine:1995ag}
\bibinfo{author}{\bibfnamefont{M.}~\bibnamefont{Dine}},
  \bibinfo{author}{\bibfnamefont{A.~E.} \bibnamefont{Nelson}},
  \bibinfo{author}{\bibfnamefont{Y.}~\bibnamefont{Nir}}, \bibnamefont{and}
  \bibinfo{author}{\bibfnamefont{Y.}~\bibnamefont{Shirman}},
  \bibinfo{journal}{Phys. Rev.} \textbf{\bibinfo{volume}{D53}},
  \bibinfo{pages}{2658} (\bibinfo{year}{1996}), \eprint{hep-ph/9507378}.

\bibitem[{\citenamefont{Giudice and Rattazzi}(1999)}]{Giudice:1998bp}
\bibinfo{author}{\bibfnamefont{G.~F.} \bibnamefont{Giudice}} \bibnamefont{and}
  \bibinfo{author}{\bibfnamefont{R.}~\bibnamefont{Rattazzi}},
  \bibinfo{journal}{Phys. Rept.} \textbf{\bibinfo{volume}{322}},
  \bibinfo{pages}{419} (\bibinfo{year}{1999}), \eprint{hep-ph/9801271}.

\bibitem[{\citenamefont{Abdallah et~al.}(2004)}]{Abdallah:2003xe}
\bibinfo{author}{\bibfnamefont{J.}~\bibnamefont{Abdallah}} \bibnamefont{et~al.}
  (\bibinfo{collaboration}{DELPHI}), \bibinfo{journal}{Eur. Phys. J.}
  \textbf{\bibinfo{volume}{C31}}, \bibinfo{pages}{421} (\bibinfo{year}{2004}),
  \eprint{hep-ex/0311019}.

\bibitem[{\citenamefont{Abbiendi et~al.}(2004)}]{Abbiendi:2003sc}
\bibinfo{author}{\bibfnamefont{G.}~\bibnamefont{Abbiendi}} \bibnamefont{et~al.}
  (\bibinfo{collaboration}{OPAL}), \bibinfo{journal}{Eur. Phys. J.}
  \textbf{\bibinfo{volume}{C35}}, \bibinfo{pages}{1} (\bibinfo{year}{2004}),
  \eprint{hep-ex/0401026}.

\bibitem[{\citenamefont{Dvali et~al.}(1996)\citenamefont{Dvali, Giudice, and
  Pomarol}}]{Dvali:1996cu}
\bibinfo{author}{\bibfnamefont{G.~R.} \bibnamefont{Dvali}},
  \bibinfo{author}{\bibfnamefont{G.~F.} \bibnamefont{Giudice}},
  \bibnamefont{and} \bibinfo{author}{\bibfnamefont{A.}~\bibnamefont{Pomarol}},
  \bibinfo{journal}{Nucl. Phys.} \textbf{\bibinfo{volume}{B478}},
  \bibinfo{pages}{31} (\bibinfo{year}{1996}), \eprint{hep-ph/9603238}.

\bibitem[{\citenamefont{Delgado et~al.}(2007)\citenamefont{Delgado, Giudice,
  and Slavich}}]{Delgado:2007rz}
\bibinfo{author}{\bibfnamefont{A.}~\bibnamefont{Delgado}},
  \bibinfo{author}{\bibfnamefont{G.~F.} \bibnamefont{Giudice}},
  \bibnamefont{and} \bibinfo{author}{\bibfnamefont{P.}~\bibnamefont{Slavich}},
  \bibinfo{journal}{Phys. Lett.} \textbf{\bibinfo{volume}{B653}},
  \bibinfo{pages}{424} (\bibinfo{year}{2007}), \eprint{arXiv:0706.3873 [hep-ph]}.

\bibitem[{\citenamefont{Giudice et~al.}(2008)\citenamefont{Giudice, Kim, and
  Rattazzi}}]{Giudice:2007ca}
\bibinfo{author}{\bibfnamefont{G.~F.} \bibnamefont{Giudice}},
  \bibinfo{author}{\bibfnamefont{H.~D.} \bibnamefont{Kim}}, \bibnamefont{and}
  \bibinfo{author}{\bibfnamefont{R.}~\bibnamefont{Rattazzi}},
  \bibinfo{journal}{Phys. Lett.} \textbf{\bibinfo{volume}{B660}},
  \bibinfo{pages}{545} (\bibinfo{year}{2008}), \eprint{arXiv:0711.4448  [hep-ph]}.

\bibitem[{\citenamefont{Chacko and Ponton}(2002)}]{Chacko:2001km}
\bibinfo{author}{\bibfnamefont{Z.}~\bibnamefont{Chacko}} \bibnamefont{and}
  \bibinfo{author}{\bibfnamefont{E.}~\bibnamefont{Ponton}},
  \bibinfo{journal}{Phys. Rev.} \textbf{\bibinfo{volume}{D66}},
  \bibinfo{pages}{095004} (\bibinfo{year}{2002}), \eprint{hep-ph/0112190  [hep-ph]}.

\bibitem[{\citenamefont{Murayama et~al.}(2008)\citenamefont{Murayama, Nomura,
  and Poland}}]{Murayama:2007ge}
\bibinfo{author}{\bibfnamefont{H.}~\bibnamefont{Murayama}},
  \bibinfo{author}{\bibfnamefont{Y.}~\bibnamefont{Nomura}}, \bibnamefont{and}
  \bibinfo{author}{\bibfnamefont{D.}~\bibnamefont{Poland}},
  \bibinfo{journal}{Phys. Rev.} \textbf{\bibinfo{volume}{D77}},
  \bibinfo{pages}{015005} (\bibinfo{year}{2008}), \eprint{arXiv:0709.0775  [hep-ph]}.

\bibitem[{\citenamefont{Roy and Schmaltz}(2008)}]{Roy:2007nz}
\bibinfo{author}{\bibfnamefont{T.~S.} \bibnamefont{Roy}} \bibnamefont{and}
  \bibinfo{author}{\bibfnamefont{M.}~\bibnamefont{Schmaltz}},
  \bibinfo{journal}{Phys. Rev.} \textbf{\bibinfo{volume}{D77}},
  \bibinfo{pages}{095008} (\bibinfo{year}{2008}), \eprint{arXiv:0708.3593  [hep-ph]}.

\bibitem[{\citenamefont{Schael et~al.}(2006)}]{Schael:2006cr}
\bibinfo{author}{\bibfnamefont{S.}~\bibnamefont{Schael}} \bibnamefont{et~al.}
  (\bibinfo{collaboration}{ALEPH}), \bibinfo{journal}{Eur. Phys. J.}
  \textbf{\bibinfo{volume}{C47}}, \bibinfo{pages}{547} (\bibinfo{year}{2006}),
  \eprint{hep-ex/0602042}.

\bibitem[{\citenamefont{Haber and Hempfling}(1991)}]{Haber:1990aw}
\bibinfo{author}{\bibfnamefont{H.~E.} \bibnamefont{Haber}} \bibnamefont{and}
  \bibinfo{author}{\bibfnamefont{R.}~\bibnamefont{Hempfling}},
  \bibinfo{journal}{Phys. Rev. Lett.} \textbf{\bibinfo{volume}{66}},
  \bibinfo{pages}{1815} (\bibinfo{year}{1991}).

\bibitem[{\citenamefont{Dine}(2009)}]{Dine:2009gy}
\bibinfo{author}{\bibfnamefont{M.}~\bibnamefont{Dine}} (\bibinfo{year}{2009}),
  \eprint{arXiv:0901.1713  [hep-ph]}.

\bibitem[{\citenamefont{{Yao} et~al.}(2006)\citenamefont{{Yao}, {Amsler},
  {Asner}, {Barnett}, {Beringer}, {Burchat}, {Carone}, {Caso}, {Dahl},
  {D'Ambrosio} et~al.}}]{PDBook}
\bibinfo{author}{\bibfnamefont{W.-M.} \bibnamefont{{Yao}}},
  \bibinfo{author}{\bibfnamefont{C.}~\bibnamefont{{Amsler}}},
  \bibinfo{author}{\bibfnamefont{D.}~\bibnamefont{{Asner}}},
  \bibinfo{author}{\bibfnamefont{R.}~\bibnamefont{{Barnett}}},
  \bibinfo{author}{\bibfnamefont{J.}~\bibnamefont{{Beringer}}},
  \bibinfo{author}{\bibfnamefont{P.}~\bibnamefont{{Burchat}}},
  \bibinfo{author}{\bibfnamefont{C.}~\bibnamefont{{Carone}}},
  \bibinfo{author}{\bibfnamefont{C.}~\bibnamefont{{Caso}}},
  \bibinfo{author}{\bibfnamefont{O.}~\bibnamefont{{Dahl}}},
  \bibinfo{author}{\bibfnamefont{G.}~\bibnamefont{{D'Ambrosio}}},
  \bibnamefont{et~al.}, \bibinfo{journal}{{Journal of Physics G}}
  \textbf{\bibinfo{volume}{33}}, \bibinfo{pages}{1+} (\bibinfo{year}{2006}),
  \urlprefix\url{http://pdg.lbl.gov}.

\bibitem[{\citenamefont{Izawa et~al.}(1997)\citenamefont{Izawa, Nomura, Tobe,
  and Yanagida}}]{Izawa:1997gs}
\bibinfo{author}{\bibfnamefont{K.~I.} \bibnamefont{Izawa}},
  \bibinfo{author}{\bibfnamefont{Y.}~\bibnamefont{Nomura}},
  \bibinfo{author}{\bibfnamefont{K.}~\bibnamefont{Tobe}}, \bibnamefont{and}
  \bibinfo{author}{\bibfnamefont{T.}~\bibnamefont{Yanagida}},
  \bibinfo{journal}{Phys. Rev.} \textbf{\bibinfo{volume}{D56}},
  \bibinfo{pages}{2886} (\bibinfo{year}{1997}), \eprint{hep-ph/9705228}.

\bibitem[{\citenamefont{Dimopoulos et~al.}(1997)\citenamefont{Dimopoulos,
  Thomas, and Wells}}]{Dimopoulos:1996yq}
\bibinfo{author}{\bibfnamefont{S.}~\bibnamefont{Dimopoulos}},
  \bibinfo{author}{\bibfnamefont{S.~D.} \bibnamefont{Thomas}},
  \bibnamefont{and} \bibinfo{author}{\bibfnamefont{J.~D.} \bibnamefont{Wells}},
  \bibinfo{journal}{Nucl. Phys.} \textbf{\bibinfo{volume}{B488}},
  \bibinfo{pages}{39} (\bibinfo{year}{1997}), \eprint{hep-ph/9609434}.
  
  \bibitem[{\citenamefont{Martin}(1997)}]{Martin:1996zb}
\bibinfo{author}{\bibfnamefont{S.~P.} \bibnamefont{Martin}},
  \bibinfo{journal}{Phys. Rev.} \textbf{\bibinfo{volume}{D55}},
  \bibinfo{pages}{3177} (\bibinfo{year}{1997}), \eprint{hep-ph/9608224}.
  
  \bibitem[{\citenamefont{Agashe and Graesser}(1997)}]{Agashe:1997kn}
\bibinfo{author}{\bibfnamefont{K.}~\bibnamefont{Agashe}} \bibnamefont{and}
  \bibinfo{author}{\bibfnamefont{M.}~\bibnamefont{Graesser}},
  \bibinfo{journal}{Nucl. Phys.} \textbf{\bibinfo{volume}{B507}},
  \bibinfo{pages}{3} (\bibinfo{year}{1997}), \eprint{hep-ph/9704206}.
  
  \bibitem[{\citenamefont{Dine and Mason}(2008)}]{Dine:2007dz}
\bibinfo{author}{\bibfnamefont{M.}~\bibnamefont{Dine}} \bibnamefont{and}
  \bibinfo{author}{\bibfnamefont{J.~D.} \bibnamefont{Mason}},
  \bibinfo{journal}{Phys. Rev.} \textbf{\bibinfo{volume}{D78}},
  \bibinfo{pages}{055013} (\bibinfo{year}{2008}), \eprint{arXiv:0712.1355 [hep-ph]}.

\bibitem[{\citenamefont{Cheung et~al.}(2008)\citenamefont{Cheung, Fitzpatrick,
  and Shih}}]{Cheung:2007es}
\bibinfo{author}{\bibfnamefont{C.}~\bibnamefont{Cheung}},
  \bibinfo{author}{\bibfnamefont{A.~L.} \bibnamefont{Fitzpatrick}},
  \bibnamefont{and} \bibinfo{author}{\bibfnamefont{D.}~\bibnamefont{Shih}},
  \bibinfo{journal}{JHEP} \textbf{\bibinfo{volume}{07}}, \bibinfo{pages}{054}
  (\bibinfo{year}{2008}), \eprint{arXiv:0710.3585  [hep-ph]}.

\bibitem[{\citenamefont{Buican et~al.}(2008)\citenamefont{Buican, Meade,
  Seiberg, and Shih}}]{Buican:2008ws}
\bibinfo{author}{\bibfnamefont{M.}~\bibnamefont{Buican}},
  \bibinfo{author}{\bibfnamefont{P.}~\bibnamefont{Meade}},
  \bibinfo{author}{\bibfnamefont{N.}~\bibnamefont{Seiberg}}, \bibnamefont{and}
  \bibinfo{author}{\bibfnamefont{D.}~\bibnamefont{Shih}}
  (\bibinfo{year}{2008}), \eprint{arXiv:0812.3668  [hep-ph]}.

\bibitem[{\citenamefont{Abel et~al.}(2009)\citenamefont{Abel, Jaeckel, Khoze,
  and Matos}}]{Abel:2008gv}
\bibinfo{author}{\bibfnamefont{S.}~\bibnamefont{Abel}},
  \bibinfo{author}{\bibfnamefont{J.}~\bibnamefont{Jaeckel}},
  \bibinfo{author}{\bibfnamefont{V.~V.} \bibnamefont{Khoze}}, \bibnamefont{and}
  \bibinfo{author}{\bibfnamefont{L.}~\bibnamefont{Matos}},
  \bibinfo{journal}{JHEP} \textbf{\bibinfo{volume}{03}}, \bibinfo{pages}{017}
  (\bibinfo{year}{2009}), \eprint{arXiv:0812.3119  [hep-ph]}.

\bibitem[{\citenamefont{Meade et~al.}(2008)\citenamefont{Meade, Seiberg, and
  Shih}}]{Meade:2008wd}
\bibinfo{author}{\bibfnamefont{P.}~\bibnamefont{Meade}},
  \bibinfo{author}{\bibfnamefont{N.}~\bibnamefont{Seiberg}}, \bibnamefont{and}
  \bibinfo{author}{\bibfnamefont{D.}~\bibnamefont{Shih}}
  (\bibinfo{year}{2008}), \eprint{arXiv:0801.3278  [hep-ph]}.

\bibitem[{\citenamefont{Carpenter et~al.}(2009)\citenamefont{Carpenter, Dine,
  Festuccia, and Mason}}]{Carpenter:2008wi}
\bibinfo{author}{\bibfnamefont{L.~M.} \bibnamefont{Carpenter}},
  \bibinfo{author}{\bibfnamefont{M.}~\bibnamefont{Dine}},
  \bibinfo{author}{\bibfnamefont{G.}~\bibnamefont{Festuccia}},
  \bibnamefont{and} \bibinfo{author}{\bibfnamefont{J.~D.} \bibnamefont{Mason}},
  \bibinfo{journal}{Phys. Rev.} \textbf{\bibinfo{volume}{D79}},
  \bibinfo{pages}{035002} (\bibinfo{year}{2009}), \eprint{arXiv:0805.2944  [hep-ph]}.
  
  \bibitem[{\citenamefont{Carpenter}(2008{\natexlab{a}})}]{Carpenter:2008rj}
\bibinfo{author}{\bibfnamefont{L.~M.} \bibnamefont{Carpenter}}
  (\bibinfo{year}{2008}{\natexlab{a}}), \eprint{arXiv:0809.0026 [hep-ph]}.

\bibitem[{\citenamefont{Komargodski and Seiberg}(2008)}]{Komargodski:2008ax}
\bibinfo{author}{\bibfnamefont{Z.}~\bibnamefont{Komargodski}} \bibnamefont{and}
  \bibinfo{author}{\bibfnamefont{N.}~\bibnamefont{Seiberg}}
  (\bibinfo{year}{2008}), \eprint{arXiv:0812.3900  [hep-ph]}.

\bibitem[{\citenamefont{Csaki et~al.}(2008)\citenamefont{Csaki, Falkowski,
  Nomura, and Volansky}}]{Csaki:2008sr}
\bibinfo{author}{\bibfnamefont{C.}~\bibnamefont{Csaki}},
  \bibinfo{author}{\bibfnamefont{A.}~\bibnamefont{Falkowski}},
  \bibinfo{author}{\bibfnamefont{Y.}~\bibnamefont{Nomura}}, \bibnamefont{and}
  \bibinfo{author}{\bibfnamefont{T.}~\bibnamefont{Volansky}}
  (\bibinfo{year}{2008}), \eprint{arXiv:0809.4492  [hep-ph]}.

\bibitem[{\citenamefont{Liu and Wagner}(2008)}]{Liu:2008pa}
\bibinfo{author}{\bibfnamefont{T.}~\bibnamefont{Liu}} \bibnamefont{and}
  \bibinfo{author}{\bibfnamefont{C.~E.~M.} \bibnamefont{Wagner}},
  \bibinfo{journal}{JHEP} \textbf{\bibinfo{volume}{06}}, \bibinfo{pages}{073}
  (\bibinfo{year}{2008}), \eprint{arXiv:0803.2895  [hep-ph]}.

\bibitem[{\citenamefont{Dermisek and Gunion}(2005)}]{Dermisek:2005ar}
\bibinfo{author}{\bibfnamefont{R.}~\bibnamefont{Dermisek}} \bibnamefont{and}
  \bibinfo{author}{\bibfnamefont{J.~F.} \bibnamefont{Gunion}},
  \bibinfo{journal}{Phys. Rev. Lett.} \textbf{\bibinfo{volume}{95}},
  \bibinfo{pages}{041801} (\bibinfo{year}{2005}), \eprint{hep-ph/0502105}.
  
    \bibitem[{\citenamefont{Chang et~al.}(2006)\citenamefont{Chang, Fox, and
  Weiner}}]{Chang:2005ht}
\bibinfo{author}{\bibfnamefont{S.}~\bibnamefont{Chang}},
  \bibinfo{author}{\bibfnamefont{P.~J.} \bibnamefont{Fox}}, \bibnamefont{and}
  \bibinfo{author}{\bibfnamefont{N.}~\bibnamefont{Weiner}},
  \bibinfo{journal}{JHEP} \textbf{\bibinfo{volume}{08}}, \bibinfo{pages}{068}
  (\bibinfo{year}{2006}), \eprint{hep-ph/0511250}.

\bibitem[{\citenamefont{Dermisek and Gunion}(2007)}]{Dermisek:2006wr}
\bibinfo{author}{\bibfnamefont{R.}~\bibnamefont{Dermisek}} \bibnamefont{and}
  \bibinfo{author}{\bibfnamefont{J.~F.} \bibnamefont{Gunion}},
  \bibinfo{journal}{Phys. Rev.} \textbf{\bibinfo{volume}{D75}},
  \bibinfo{pages}{075019} (\bibinfo{year}{2007}), \eprint{hep-ph/0611142}.

\bibitem[{\citenamefont{Chang et~al.}(2008)\citenamefont{Chang, Dermisek,
  Gunion, and Weiner}}]{Chang:2008cw}
\bibinfo{author}{\bibfnamefont{S.}~\bibnamefont{Chang}},
  \bibinfo{author}{\bibfnamefont{R.}~\bibnamefont{Dermisek}},
  \bibinfo{author}{\bibfnamefont{J.~F.} \bibnamefont{Gunion}},
  \bibnamefont{and} \bibinfo{author}{\bibfnamefont{N.}~\bibnamefont{Weiner}},
  \bibinfo{journal}{Ann. Rev. Nucl. Part. Sci.} \textbf{\bibinfo{volume}{58}},
  \bibinfo{pages}{75} (\bibinfo{year}{2008}), \eprint{arXiv:0801.4554  [hep-ph]}.

\bibitem[{\citenamefont{Haber and Hempfling}(1993)}]{Haber:1993an}
\bibinfo{author}{\bibfnamefont{H.~E.} \bibnamefont{Haber}} \bibnamefont{and}
  \bibinfo{author}{\bibfnamefont{R.}~\bibnamefont{Hempfling}},
  \bibinfo{journal}{Phys. Rev.} \textbf{\bibinfo{volume}{D48}},
  \bibinfo{pages}{4280} (\bibinfo{year}{1993}), \eprint{hep-ph/9307201}.
  
  \bibitem[{\citenamefont{Brignole et~al.}(2003)\citenamefont{Brignole, Casas,
  Espinosa, and Navarro}}]{Brignole:2003cm}
\bibinfo{author}{\bibfnamefont{A.}~\bibnamefont{Brignole}},
  \bibinfo{author}{\bibfnamefont{J.~A.} \bibnamefont{Casas}},
  \bibinfo{author}{\bibfnamefont{J.~R.} \bibnamefont{Espinosa}},
  \bibnamefont{and} \bibinfo{author}{\bibfnamefont{I.}~\bibnamefont{Navarro}},
  \bibinfo{journal}{Nucl. Phys.} \textbf{\bibinfo{volume}{B666}},
  \bibinfo{pages}{105} (\bibinfo{year}{2003}), \eprint{hep-ph/0301121}.

\bibitem[{\citenamefont{Dine et~al.}(2007)\citenamefont{Dine, Seiberg, and
  Thomas}}]{Dine:2007xi}
\bibinfo{author}{\bibfnamefont{M.}~\bibnamefont{Dine}},
  \bibinfo{author}{\bibfnamefont{N.}~\bibnamefont{Seiberg}}, \bibnamefont{and}
  \bibinfo{author}{\bibfnamefont{S.}~\bibnamefont{Thomas}},
  \bibinfo{journal}{Phys. Rev.} \textbf{\bibinfo{volume}{D76}},
  \bibinfo{pages}{095004} (\bibinfo{year}{2007}), \eprint{arXiv:0707.0005  [hep-ph]}.

\bibitem[{\citenamefont{Ellis et~al.}(1989)\citenamefont{Ellis, Gunion, Haber,
  Roszkowski, and Zwirner}}]{Ellis:1988er}
\bibinfo{author}{\bibfnamefont{J.~R.} \bibnamefont{Ellis}},
  \bibinfo{author}{\bibfnamefont{J.~F.} \bibnamefont{Gunion}},
  \bibinfo{author}{\bibfnamefont{H.~E.} \bibnamefont{Haber}},
  \bibinfo{author}{\bibfnamefont{L.}~\bibnamefont{Roszkowski}},
  \bibnamefont{and} \bibinfo{author}{\bibfnamefont{F.}~\bibnamefont{Zwirner}},
  \bibinfo{journal}{Phys. Rev.} \textbf{\bibinfo{volume}{D39}},
  \bibinfo{pages}{844} (\bibinfo{year}{1989}).

\bibitem[{\citenamefont{Espinosa and
  Quiros}(1992{\natexlab{a}})}]{Espinosa:1991wt}
\bibinfo{author}{\bibfnamefont{J.~R.} \bibnamefont{Espinosa}} \bibnamefont{and}
  \bibinfo{author}{\bibfnamefont{M.}~\bibnamefont{Quiros}},
  \bibinfo{journal}{Nucl. Phys.} \textbf{\bibinfo{volume}{B384}},
  \bibinfo{pages}{113} (\bibinfo{year}{1992}{\natexlab{a}}).

\bibitem[{\citenamefont{Espinosa and
  Quiros}(1992{\natexlab{b}})}]{Espinosa:1991gr}
\bibinfo{author}{\bibfnamefont{J.~R.} \bibnamefont{Espinosa}} \bibnamefont{and}
  \bibinfo{author}{\bibfnamefont{M.}~\bibnamefont{Quiros}},
  \bibinfo{journal}{Phys. Lett.} \textbf{\bibinfo{volume}{B279}},
  \bibinfo{pages}{92} (\bibinfo{year}{1992}{\natexlab{b}}).

\bibitem[{\citenamefont{Nomura et~al.}(2006)\citenamefont{Nomura, Poland, and
  Tweedie}}]{Nomura:2005rk}
\bibinfo{author}{\bibfnamefont{Y.}~\bibnamefont{Nomura}},
  \bibinfo{author}{\bibfnamefont{D.}~\bibnamefont{Poland}}, \bibnamefont{and}
  \bibinfo{author}{\bibfnamefont{B.}~\bibnamefont{Tweedie}},
  \bibinfo{journal}{Phys. Lett.} \textbf{\bibinfo{volume}{B633}},
  \bibinfo{pages}{573} (\bibinfo{year}{2006}), \eprint{hep-ph/0509244}.

\bibitem[{\citenamefont{Batra et~al.}(2004)\citenamefont{Batra, Delgado,
  Kaplan, and Tait}}]{Batra:2004vc}
\bibinfo{author}{\bibfnamefont{P.}~\bibnamefont{Batra}},
  \bibinfo{author}{\bibfnamefont{A.}~\bibnamefont{Delgado}},
  \bibinfo{author}{\bibfnamefont{D.~E.} \bibnamefont{Kaplan}},
  \bibnamefont{and} \bibinfo{author}{\bibfnamefont{T.~M.~P.}
  \bibnamefont{Tait}}, \bibinfo{journal}{JHEP} \textbf{\bibinfo{volume}{06}},
  \bibinfo{pages}{032} (\bibinfo{year}{2004}), \eprint{hep-ph/0404251}.

\bibitem[{\citenamefont{Barbieri et~al.}(2008)\citenamefont{Barbieri, Hall,
  Papaioannou, Pappadopulo, and Rychkov}}]{Barbieri:2007tu}
\bibinfo{author}{\bibfnamefont{R.}~\bibnamefont{Barbieri}},
  \bibinfo{author}{\bibfnamefont{L.~J.} \bibnamefont{Hall}},
  \bibinfo{author}{\bibfnamefont{A.~Y.} \bibnamefont{Papaioannou}},
  \bibinfo{author}{\bibfnamefont{D.}~\bibnamefont{Pappadopulo}},
  \bibnamefont{and} \bibinfo{author}{\bibfnamefont{V.~S.}
  \bibnamefont{Rychkov}}, \bibinfo{journal}{JHEP}
  \textbf{\bibinfo{volume}{03}}, \bibinfo{pages}{005} (\bibinfo{year}{2008}),
  \eprint{arXiv:0712.2903  [hep-ph]}.

\bibitem[{\citenamefont{Cavicchia}(2008)}]{Cavicchia:2008fn}
\bibinfo{author}{\bibfnamefont{L.}~\bibnamefont{Cavicchia}}
  (\bibinfo{year}{2008}), \eprint{arXiv:0807.3921 [hep-ph]}.

\bibitem[{\citenamefont{Alwall et~al.}(2008)\citenamefont{Alwall, Le, Lisanti,
  and Wacker}}]{Alwall:2008zz}
\bibinfo{author}{\bibfnamefont{J.}~\bibnamefont{Alwall}},
  \bibinfo{author}{\bibfnamefont{M.~P.} \bibnamefont{Le}},
  \bibinfo{author}{\bibfnamefont{M.}~\bibnamefont{Lisanti}}, \bibnamefont{and}
  \bibinfo{author}{\bibfnamefont{J.~G.} \bibnamefont{Wacker}},
  \bibinfo{journal}{Int. J. Mod. Phys.} \textbf{\bibinfo{volume}{A23}},
  \bibinfo{pages}{4637} (\bibinfo{year}{2008}).

\bibitem[{\citenamefont{Carpenter}(2008)}]{Carpenter:2008he}
\bibinfo{author}{\bibfnamefont{L.~M.} \bibnamefont{Carpenter}}
  (\bibinfo{year}{2008}), \eprint{arXiv:0812.2051  [hep-ph]}.

\bibitem[{\citenamefont{Rajaraman et~al.}(2009)\citenamefont{Rajaraman,
  Shirman, Smidt, and Yu}}]{Rajaraman:2009ga}
\bibinfo{author}{\bibfnamefont{A.}~\bibnamefont{Rajaraman}},
  \bibinfo{author}{\bibfnamefont{Y.}~\bibnamefont{Shirman}},
  \bibinfo{author}{\bibfnamefont{J.}~\bibnamefont{Smidt}}, \bibnamefont{and}
  \bibinfo{author}{\bibfnamefont{F.}~\bibnamefont{Yu}} (\bibinfo{year}{2009}),
  \eprint{arXiv:0903.0668  [hep-ph]}.

\bibitem[{\citenamefont{de~Gouvea et~al.}(1998)\citenamefont{de~Gouvea,
  Friedland, and Murayama}}]{deGouvea:1997cx}
\bibinfo{author}{\bibfnamefont{A.}~\bibnamefont{de~Gouvea}},
  \bibinfo{author}{\bibfnamefont{A.}~\bibnamefont{Friedland}},
  \bibnamefont{and} \bibinfo{author}{\bibfnamefont{H.}~\bibnamefont{Murayama}},
  \bibinfo{journal}{Phys. Rev.} \textbf{\bibinfo{volume}{D57}},
  \bibinfo{pages}{5676} (\bibinfo{year}{1998}), \eprint{hep-ph/9711264}.

\bibitem[{\citenamefont{Han et~al.}(1999)\citenamefont{Han, Marfatia, and
  Zhang}}]{Han:1999jc}
\bibinfo{author}{\bibfnamefont{T.}~\bibnamefont{Han}},
  \bibinfo{author}{\bibfnamefont{D.}~\bibnamefont{Marfatia}}, \bibnamefont{and}
  \bibinfo{author}{\bibfnamefont{R.-J.} \bibnamefont{Zhang}},
  \bibinfo{journal}{Phys. Rev.} \textbf{\bibinfo{volume}{D61}},
  \bibinfo{pages}{013007} (\bibinfo{year}{1999}), \eprint{hep-ph/9906508}.

\bibitem[{\citenamefont{Ellwanger et~al.}(2008)\citenamefont{Ellwanger,
  Jean-Louis, and Teixeira}}]{Ellwanger:2008py}
\bibinfo{author}{\bibfnamefont{U.}~\bibnamefont{Ellwanger}},
  \bibinfo{author}{\bibfnamefont{C.~C.} \bibnamefont{Jean-Louis}},
  \bibnamefont{and} \bibinfo{author}{\bibfnamefont{A.~M.}
  \bibnamefont{Teixeira}}, \bibinfo{journal}{JHEP}
  \textbf{\bibinfo{volume}{05}}, \bibinfo{pages}{044} (\bibinfo{year}{2008}),
  \eprint{arXiv:0803.2962  [hep-ph]}.




\end{thebibliography}
\end{document}